\documentclass[twocolumn,english,aps,pra,superscriptaddress,floatfix]{revtex4}
\usepackage[T1]{fontenc}
\usepackage[latin9]{inputenc}
\usepackage{color}
\usepackage{bm}
\usepackage{xcolor}
\usepackage{amsmath}
\usepackage{amssymb}
\usepackage{graphicx}
\usepackage{natbib}
\usepackage{braket}
\usepackage{psfrag}
\usepackage{tikz}
\usepackage[normalem]{ulem}
\usepackage{qcircuit}

\begin{document}


\title{Quantum Risk Analysis}

\author{Stefan Woerner}
\email{wor@zurich.ibm.com}
\author{Daniel J.~Egger}%
\affiliation{%
IBM Research -- Zurich
}%


\date{\today}

\begin{abstract}
We present a quantum algorithm that analyzes risk more efficiently than Monte Carlo simulations traditionally used on classical computers. We employ quantum amplitude estimation to evaluate risk measures such as Value at Risk and Conditional Value at Risk on a gate-based quantum computer.
Additionally, we show how to implement this algorithm and how to trade off the convergence rate of the algorithm and the circuit depth. The shortest possible circuit depth - growing polynomially in the number of qubits representing the uncertainty - leads to a convergence rate of $O(M^{-2/3})$. This is already faster than classical Monte Carlo simulations which converge at a rate of $O(M^{-1/2})$. If we allow the circuit depth to grow faster, but still polynomially, the convergence rate quickly approaches the optimum of $O(M^{-1})$.
Thus, for slowly increasing circuit depths our algorithm provides a near quadratic speed-up compared to Monte Carlo methods.
We demonstrate our algorithm using two toy models. In the first model we use real hardware, such as the IBM Q Experience, to measure the financial risk in a Treasury-bill (T-bill) faced by a possible interest rate increase. In the second model, we simulate our algorithm to illustrate how a quantum computer can determine financial risk for a two-asset portfolio made up of Government debt with different maturity dates. Both models confirm the improved convergence rate over Monte Carlo methods. Using simulations, we also evaluate the impact of cross-talk and energy relaxation errors.
\end{abstract}

\maketitle


\section{\label{sec:introduction} Introduction}

Risk management plays a central role in the financial system. Value at risk (VaR) \citep{Glasserman2000}, a quantile of the loss distribution, is a widely used risk metric. Examples of use cases include the Basel III regulations under which banks are required to perform stress tests using VaR \cite{BaselIII} and the calculation of haircuts applied to collateral used in security settelement systems \cite{Garcia2006}. A second important risk metric is conditional value at risk (CVaR, sometimes also called expected shortfall), defined as the expected loss for losses greater than VaR. By contrast to VaR, CVaR is more sensitive to extreme events in the tail of the loss distribution.

Monte Carlo simulations are the method of choice to determine VaR and CVaR of a portfolio \citep{Glasserman2000}. They are done by building a model of the portfolio assets and computing the aggregated value for $M$ different realizations of the model input parameters. VaR calculations are computationally intensive since the width of the confidence interval scales as $O\left(M^{-\frac{1}{2}}\right)$. Many different runs are needed to achieve a representative distribution of the portfolio value.

Quantum computers process information using the laws of quantum mechanics \cite{Nielsen2010}. This has opened up novel ways of addressing some problems, e.g. in quantum chemistry \citep{Peruzzo2014}, optimization \citep{Moll2017}, or machine learning \citep{Biamonte2017}. Amplitude estimation is a quantum algorithm used to estimate an unknown parameter and converges as $O\left(M^{-1}\right)$, which is a quadratic speed-up over classical algorithms like Monte Carlo \cite{Brassard2000}. It has already been shown how amplitude estimation can be used to price options with the Black-Scholes model \cite{BlackScholes, Rebentrost2018}.

In Section \ref{sec:evaluation} of this paper we show how to use amplitude estimation to calculate expectation, variance, VaR and CVaR of random distributions. 
Section \ref{sec:mapping} discusses how to construct the corresponding quantum circuits.
In Sections \ref{sec:1_asset_problem} and \ref{sec:2_asset_problem} we show how to apply our algorithm to portfolios made up of debt issued by the United States Treasury (US Treasury), which as of December 2016 had 14.5 trillion USD in outstanding marketable debt held by the public \cite{PublicDebt201612}. 
This debt is an actively traded asset class with typical daily volumes close to 500 billion USD \cite{TreasuryVolume2016} and is regarded as high quality collateral \cite{TreasuryCollateral}.
Additionally, government debt typically lacks some of the more complex features that other types of fixed-income securities have.
These features make US Treasuries a highly relevant asset class to study whilst allowing us to use simple models to illustrate our algorithm. 
In Section \ref{sec:1_asset_problem} we introduce a very simple portfolio made up of one T-Bill analyzed on a single period of a binomial tree.
We demonstrate amplitude estimation and can approximate the expected value of the T-Bill on a real quantum computer.
In Section \ref{sec:2_asset_problem} we show a more comprehensive two asset portfolio and simulate the presented algorithms assuming a perfect as well as a noisy quantum computer.
We discuss our results as well as next steps in Sec. \ref{sec:conclusion}.

%

\section{\label{sec:evaluation} Quantum Risk Analysis}

In this section, we introduce \emph{amplitude estimation} and explain how it can be used to estimate properties of random distributions such as risk measures.

\begin{figure}[ht]
	\centering
		\includegraphics[width=.45\textwidth]{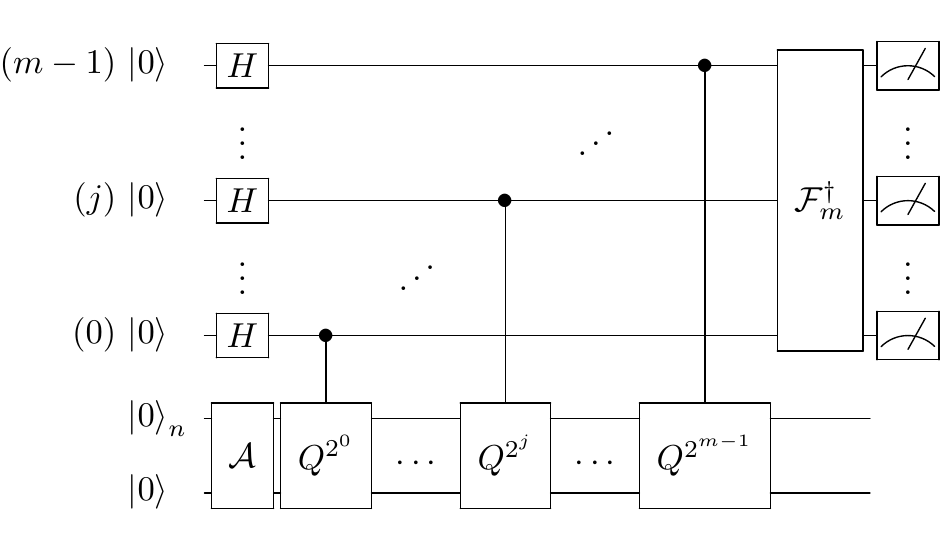}
	\caption{Quantum circuit for amplitude estimation as introduced in \cite{Brassard2000}. $H$ is the Hadamard gate and $\mathcal{F}_m^{\dagger}$ denotes the inverse Quantum Fourier Transform on $m$ qubits.}
	\label{fig:amplitude_estimation_circuit}
\end{figure}

Suppose a unitary operator $\mathcal{A}$ acting on a register of $(n+1)$ qubits such that $\mathcal{A} \ket{0}_{n+1} = \sqrt{1 - a}\ket{\psi_0}_n\ket{0} + \sqrt{a}\ket{\psi_1}_n\ket{1}$ for some normalized states $\ket{\psi_0}_n$ and $\ket{\psi_1}_n$, where $a \in [0, 1]$ is unknown.
Amplitude estimation allows the efficient estimation of $a$, i.e., the probability of measuring $\ket{1}$ in the last qubit \cite{Brassard2000}. 
This is done using an operator $Q$ (formally introduced in Appendix \ref{sec:q_operator}), based on $\mathcal{A}$, and Quantum Phase Estimation \citep{Kitaev1995} to approximate certain eigenvalues of $Q$.
This requires $m$ additional qubits and $M = 2^m$ applications of $Q$.
The $m$ qubits are first put into equal superposition by applying Hadamard gates.
Then, they are used to control different powers of $Q$.
And last, after an inverse Quantum Fourier Transform has been applied, their state is measured, see the circuit in Fig. \ref{fig:amplitude_estimation_circuit}. This results in an integer $y \in \{0, ..., M-1\}$, which is classically mapped to the estimator $\tilde{a} = \sin^2(y\pi/M) \in [0, 1]$.
The estimator $\tilde{a}$ satisfies
\begin{eqnarray}
| a - \tilde{a} | 
&\leq& \frac{2\sqrt{a(1-a)}\pi}{M} + \frac{\pi^2}{M^2} \\
&\leq& \frac{\pi}{M} + \frac{\pi^2}{M^2} \;=\; O\left(M^{-1}\right),
\end{eqnarray}
with probability of at least $\frac{8}{\pi^2}$.
This represents a quadratic speedup compared to the $O\left(M^{-1/2}\right)$ convergence rate of classical Monte Carlo methods \citep{Glasserman2000}.

We now explain how to use amplitude estimation to approximate the expected value of a random variable \citep{Abrams1999, Montanaro2017}.
Suppose a quantum state
\begin{equation}
\ket{\psi}_n = \sum_{i=0}^{N-1} \sqrt{p_i} \ket{i}_n, \label{eq:uncertainty_psi}
\end{equation}
where the probability of measuring the state $\ket{i}_n$ is $p_i \in [0, 1]$, with $\sum_{i=0}^{N-1} p_i = 1$, and $N = 2^n$.
The state $\ket{i}_n$ is one of the $N$ possible realizations of a bounded discrete random variable $X$, which, for instance, can represent a discretized interest rate or the value of a portfolio.

We consider a function $f: \{0, ..., N-1\} \rightarrow [0, 1]$ and a corresponding operator 
\begin{equation}
F: \ket{i}_n\ket{0} \mapsto \ket{i}_n \left( \sqrt{1 - f(i)}\ket{0} + \sqrt{f(i)} \ket{1} \right),
\label{eq:f_operator}
\end{equation}
for all $i \in \{0, ..., N-1\}$, acting on an ancilla qubit.
Applying $F$ to $\ket{\psi}_n\ket{0}$ leads to the state
\begin{eqnarray*}
\sum_{i=0}^{N-1} \sqrt{1 - f(i)} \sqrt{p_i} \ket{i}_n\ket{0} 
+ \sum_{i=0}^{N-1} \sqrt{f(i)} \sqrt{p_i} \ket{i}_n\ket{1}.
\end{eqnarray*}
Now we can use amplitude estimation to approximate the probability of measuring $\ket{1}$ in the last qubit, which equals
$\sum_{i=0}^{N-1} p_i f(i)$ and thus also $\mathbb{E}\left[f(X)\right]$.
Choosing $f(i) = i/(N-1)$ allows us to estimate $\mathbb{E}[\frac{X}{N-1}]$ and hence $\mathbb{E}[X]$.
If we choose $f(i) = i^2/(N-1)^2$ we can efficiently estimate $\mathbb{E}[X^2]$ which yields the variance $\text{Var}(X) = \mathbb{E}[X^2] - \mathbb{E}[X]^2$.
In the remainder of this section we extend this technique and show how to evaluate risk measures such as VaR and CVaR.

For a given confidence level $\alpha \in [0, 1]$, $\text{VaR}_{\alpha}(X)$ can be defined as the smallest value $x \in \{0, ..., N-1\}$ such that $\mathbb{P}[X \leq x] \geq (1 - \alpha)$. 
To find $\text{VaR}_{\alpha}(X)$ on a quantum computer, we define the function $f_l(i) = 1$ if $i \leq l$ and $f_l(i) = 0$ otherwise, where $l \in \{0, ..., N-1\}$.
Applying $F_l$, i.e. the operator corresponding to $f_l$, to $\ket{\psi}_n\ket{0}$ leads to the state
\begin{eqnarray}
\sum_{i=l+1}^{N-1} \sqrt{p_i} \ket{i}_n \ket{0} + \sum_{i=0}^{l} \sqrt{p_i} \ket{i}_n \ket{1}.
\end{eqnarray}
The probability of measuring $\ket{1}$ for the last qubit is $\sum_{i=0}^{l} p_i = \mathbb{P}[X \leq l]$.
Therefore, with a bisection search over $l$ we can find the smallest level $l_{\alpha}$ such that $\mathbb{P}[X \leq l_{\alpha}] \geq 1 - \alpha$ in at most $n$ steps.
The smallest level $l_{\alpha}$ is equal to $\text{VaR}_{\alpha}(X)$.
This allows us to estimate $\text{VaR}_{\alpha}(X)$ as before with accuracy $O\left(M^{-1}\right)$, which again is a quadratic speedup compared to classical Monte Carlo methods.

$\text{CVaR}_{\alpha}(X)$ is the conditional expectation of $X$ restricted to $\{0, ..., l_{\alpha}\}$, where we compute $l_{\alpha} = \text{VaR}_{\alpha}(X)$ as before.
To estimate CVaR we apply the operator $F$ that corresponds to the function $f(i) = \frac{i}{l_{\alpha}} \cdot f_{l_{\alpha}}(i)$ to $\ket{\psi}_n\ket{0}$, which leads to the state
\begin{eqnarray}
\left(\sum_{i=l_{\alpha}+1}^{N-1} \sqrt{p_i} \ket{i}_n + \sum_{i=0}^{l_{\alpha}} \sqrt{1 - \frac{i}{l_{\alpha}}} \sqrt{p_i} \ket{i}_n \right) \ket{0} \nonumber \\
 + \sum_{i=0}^{l_{\alpha}} \sqrt{\frac{i}{l_{\alpha}}} \sqrt{p_i} \ket{i}_n \ket{1}.
\label{eq:cvar_state}
\end{eqnarray}
The probability of measuring $\ket{1}$ for the last qubit equals $\sum_{i=0}^{l_{\alpha}} \frac{i}{l_{\alpha}} p_i$, which we approximate using amplitude estimation.
However, we know that $\sum_{i=0}^{l_{\alpha}} p_i$ does not sum up to one but to $\mathbb{P}[X \leq l_{\alpha}]$ as evaluated during the VaR estimation.
Therefore we must normalize the probability of measuring $\ket{1}$ to get
\begin{eqnarray}
\text{CVaR}_{\alpha}(X) = \frac{l_{\alpha}}{\mathbb{P}[X \leq l_{\alpha}]}\sum_{i=0}^{l_{\alpha}} \frac{i}{l_{\alpha}} p_i.
\end{eqnarray}
We also multiplied by $l_{\alpha}$, otherwise we would estimate $\text{CVaR}_{\alpha}\left(\frac{X}{l_{\alpha}}\right)$.
Even though we replace $\mathbb{P}[X \leq l_{\alpha}]$ by an estimation, the error bound on CVaR, computed in Appendix \ref{sec:cvar_error_bound}, shows that we still achieve a quadratic speed up compared to classical Monte Carlo methods.

We have shown how to calculate the expected value, variance, VaR and CVaR of $X$. However, if we are instead interested in properties of $g(X)$, for a given function $g: \{0, ..., N-1 \} \rightarrow \{0, ..., N^{\prime}-1\}$, $N^{\prime} = 2^{n^{\prime}}$, $n^{\prime} \in \mathbb{N}$, we can apply an operator $G: \ket{i}_n\ket{0}_{n^{\prime}} \mapsto \ket{i}_n\ket{g(i)}_{n^{\prime}}$  and use the previously introduced algorithms on the second register.
Alternatively, as long as we can efficiently perform the bisection search on $g(X) \leq l$ for $l \in \{0, ..., N^{\prime}-1\}$, we can spare the second register and combine $f$ and $g$ and apply all algorithms directly.

\section{\label{sec:mapping} Quantum Circuits}

In this section, we show how the algorithms discussed in Sec. \ref{sec:evaluation} can be mapped to quantum circuits.

We start with the construction of $\ket{\psi}_n$ as introduced in Eq. (\ref{eq:uncertainty_psi}), representing the probability distribution of a random variable $X$ mapped to $\{0, ..., N-1\}$.
In general, the best known upper bound for the number of gates required to create $\ket{\psi}_n$ is $O(2^n)$ \citep{Mottonen2004a}.
However, approximations with polynomial complexity in $n$ are possible for many distributions, e.g., log-concave distributions \cite{Grover2002}.
In the remainder of this section, we assume a given operator $\mathcal{R}$ such that $\mathcal{R}\ket{0}_n = \ket{\psi}_n$.

If we are interested in properties of $g(X)$, as discussed in the previous section, then, depending on $g$, we can use basic arithmetic operations to construct the operator $G$.
Numerous quantum algorithms exist for arithmetic operations \citep{Vedral1995, Draper2000, Cuccaro2004, Draper2004, Bhaskar2015} as well as tools to translate classical logic into quantum circuits \citep{Green2013, Green2013b}.
However, since the latter are not necessarily efficient, the development of new and improved algorithms is ongoing research.

Approximating $\mathbb{E}[X]$ using amplitude estimation requires the operator $F$ for $f(x) = x / (N-1)$, defined in Eq. (\ref{eq:f_operator}).
In general, representing $F$ for the expected value or for the CVaR either requires an exponential $O(2^n)$ number of gates or additional ancillas to pre-compute the (discretized) function $f$ into qubits, using quantum arithmetic, before applying the rotation \citep{Mitarai2018}. The exact number of ancillas depends on the desired accuracy of the approximation of $F$.
Another approach consists of piecewise polynomial approximations of $f$ \citep{Haner2018}. However, this also implies a significant overhead in terms of the number of ancillas and gates.
In the following, we show how to overcome these hurdles by approximating $F$ without ancillas using polynomially many gates, at the cost of a lower - but still faster than classical - rate of convergence.
Note that the operator required for estimating VaR is easier to construct and we can always achieve the optimal rate of convergence as discussed later in this section.

\begin{figure}[ht]
	\centering
		\includegraphics[width=.4\textwidth]{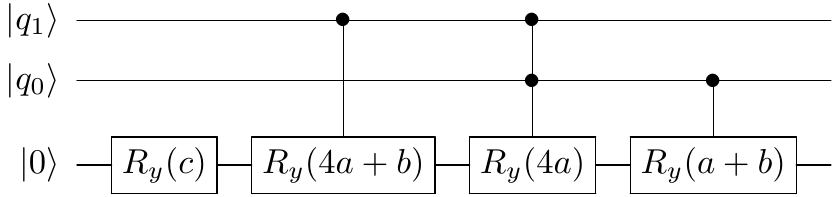}
	\caption{Quantum circuit realizing $\ket{x}_n\ket{0} \mapsto \ket{x}_n(\cos(p(x)/2)\ket{0} + \sin(p(x)/2)\ket{1})$ for $p(x) = (a x^2 + b x + c)$ and $x \in \{0, 1, 2, 3\}$. Exploiting $x = (2 q_1 + q_0)$ and $q_i^2=q_i$ leads to $p(x) = (4a+b)q_1 + 4a q_0 q_1 + (a+b) q_0 + c$, which can be directly mapped to a circuit. $R_y$ denotes a Y-rotation.}
	\label{fig:polynomial_circuit}
\end{figure}

Our contribution rests on the fact that an operator $P: \ket{x}_n \ket{0} \mapsto \ket{x}_n(\cos(p(x))\ket{0} + \sin(p(x))\ket{1})$, for a given polynomial $p(x) = \sum_{j=0}^{k} p_j x^j$ of order $k$, can be efficiently constructed using multi- controlled Y-rotations, as illustrated in Fig. \ref{fig:polynomial_circuit}.
Single qubit operations with $n-1$ control qubits can be exactly constructed, e.g., using $O(n)$ gates and $O(n)$ ancillas or $O(n^2)$ gates without any ancillas.
They can also be approximated with accuracy $\epsilon > 0$ using $O(n \log(1/\epsilon))$ gates \citep{Barenco1995}.
For simplicity, we use $O(n)$ gates and $O(n)$ ancillas.
Since the binary variable representation of $p$, illustrated in Fig. \ref{fig:polynomial_circuit}, leads to at most $n^k$ terms, the operator $P$ can be constructed using $O(n^{k+1})$ gates and $O(n)$ ancillas.

For every analytic function $f$, there exists a sequence of polynomials such that the approximation error converges exponentially fast to zero with increasing order of the polynomials \citep{Trefethen}.
Thus, for simplicity, we assume that $f$ is a polynomial of order $s$.

If we can find a polynomial $p(y)$ such that $\sin^2(p(y)) = y$, then we can set $y = f(x)$, and the previous discussion provides a way to construct the operator $F$.
Since the expected value is linear, we may choose to estimate $\mathbb{E}\left[c \left(f(X) - \frac{1}{2}\right) + \frac{1}{2}\right]$ instead of $\mathbb{E}[f(X)]$ for a parameter $c \in (0, 1]$, and then map the result back to an estimator for $\mathbb{E}[f(X)]$.
The rationale behind this choice is that $\sin^2\left(y + \frac{\pi}{4}\right) = y + \frac{1}{2} + O(y^3)$.
Thus, we want to find $p(y)$ such that $c \left(y - \frac{1}{2}\right) + \frac{1}{2}$ is sufficiently well approximated by $\sin^2\left(c p(y) + \frac{\pi}{4}\right)$.
Setting the two terms equal and solving for $p(y)$ leads to
\begin{eqnarray}
p(y) &=& \frac{1}{c}\left(\sin^{-1}\left(\sqrt{c \left(y-\frac{1}{2}\right) + \frac{1}{2}}\right) - \frac{\pi}{4} \right), \label{eq:polynomial_definition}
\end{eqnarray}
and we choose $p(y)$ as a Taylor approximation of Eq. (\ref{eq:polynomial_definition}) around $y = 1/2$.
Note that Eq. (\ref{eq:polynomial_definition}) defines an odd function around $y = 1/2$, and thus the even terms in the Taylor series equal zero.
The Taylor approximation of order $2u+1$ leads to a maximal approximation error for Eq. (\ref{eq:polynomial_definition}) of 
\begin{eqnarray}
\frac{c^{2u+3}}{(2u+3) 2^{u+1}} + O(c^{2u+5}), \label{eq:taylor_error_bound}
\end{eqnarray}
for all $y \in [0, 1]$, as shown in Appendix \ref{sec:taylor_error_bound}.

Now we consider the resulting polynomial $p(f(x))$ of order $s(2u+1)$.
The number of gates required to construct the corresponding circuit scales as $O\left(n^{s (2u+1)+1}\right)$.
The smallest scenario of interest is $s=1$ and $u=0$, i.e., both, $f$ and $p$, are linear functions, which leads to a circuit for $F$ where the number of gates scales quadratically as the number of qubits $n$ representing $\ket{\psi}_n$ grows linearly.

Thus, using amplitude estimation to estimate $\mathbb{E}\left[c (f(x)-\frac{1}{2}) + \frac{1}{2}\right]$ leads to a maximal error
\begin{align}
\frac{\pi}{M} + \frac{c^{2u+3}}{(2u+3) 2^{u+1}} + O\left(c^{2u+5} + M^{-2}\right),
\label{eq:approx_error_1}
\end{align}
where we ignore the higher order terms in the following.
Since our estimation uses $c f(x)$, we also need to analyze the scaled error $c \epsilon$, where $\epsilon > 0$ denotes the resulting estimation error for $\mathbb{E}[f(X)]$. Setting Eq. (\ref{eq:approx_error_1}) equal to $c \epsilon$ and reformulating it leads to
\begin{align}
c \epsilon - \frac{c^{2u+3}}{(2u+3) 2^{u+1}} = \frac{\pi}{M}.
\label{eq:approx_error_2}
\end{align}
Maximizing the left-hand-side with respect to $c$, i.e. minimizing the number of required samples $M$ to achieve a target error $\epsilon$, results in $c^* = \sqrt{2} \epsilon^{\frac{1}{2u+2}}$.
Plugging $c^*$ into Eq. (\ref{eq:approx_error_2}) gives
\begin{align}
\sqrt{2} \left(1 - \frac{1}{2u+3}\right) \epsilon^{ 1 + \frac{1}{2u+2} } = \frac{\pi}{M}.
\end{align}
Translating this into a rate of convergence for the estimation error $\epsilon$ with respect to the number of samples $M$ leads to $\epsilon = O\left(M^{-\frac{2u+2}{2u+3}}\right).$
For $u=0$, we get  $O\left(M^{-\frac{2}{3}}\right)$, which is already better than the classical convergence rate of $O\left(M^{-\frac{1}{2}}\right)$.
For increasing $u$, the convergence rate quickly approaches the optimal rate of $O\left(M^{-1}\right)$.

For the estimation of the expectation we exploited $\sin^2(y + \frac{\pi}{4}) \approx y + \frac{1}{2}$.
For the variance we apply the same idea but use $\sin^2(y) \approx y^2$.
We employ this approximation to estimate the value of $\mathbb{E}\left[f(X)^2\right]$ and then, together with the estimation for $\mathbb{E}\left[f(X)\right]$, we evaluate $\text{Var}\left(f(X)\right) = \mathbb{E}\left[f(X)^2\right] - \mathbb{E}\left[f(X)\right]^2$.
The resulting convergence rate is again equal to $O\left(M^{-\frac{2u+2}{2u+3}} \right)$.

The previous discussion shows how to build quantum circuits to estimate $\mathbb{E}[f(X)]$ and $\text{Var}(f(X))$ more efficiently than possible classically.
In the following, we extend this to VaR and CVaR.

Suppose the state $\ket{\psi}_n$ corresponding to the random variable $X$ on $\{0, ..., N-1\}$ and a fixed $l \in \{0, ..., N-1\}$.
To estimate VaR, we need an operator $F_l$ that maps $\ket{x}_n\ket{0}$ to $\ket{x}_n\ket{1}$ if $x \leq l$ and to $\ket{x}_n\ket{0}$ otherwise, for all $x \in \{0, ..., N-1\}$.
Then, for the fixed $l$, amplitude estimation can be used to approximate $\mathbb{P}[X \leq l]$, as shown in Eq. (\ref{eq:cvar_state}).
With $(n+1)$ ancillas, adder-circuits can be used to construct $F_l$ using $O(n)$ gates \citep{Cuccaro2004}, and the resulting convergence rate is $O\left(M^{-1}\right)$.
For a given level $\alpha$, a bisection search can find the smallest $l_{\alpha}$ such that $\mathbb{P}[X \leq l_{\alpha}] \geq \alpha$ in at most $n$ steps, and we get $l_{\alpha} = \text{VaR}_{\alpha}(X)$.

To estimate the CVaR, we apply the circuit $F_l$ for $l_{\alpha}$ to an ancilla qubit and use this ancilla qubit as a control for the operator $F$ used to estimate the expected value, but with a different normalization, as shown in Eq. (\ref{eq:cvar_state}).
Based on the previous discussion, it follows that amplitude estimation can then be used to approximate $\text{CVaR}_{\alpha}(X)$ with the same trade-off between circuit depth and convergence rate as for the expected value.

\section{\label{sec:1_asset_problem} T-Bill on a single period binomial tree}

Our first model consists of a zero coupon bond discounted at an interest rate $r$.
We seek to find the value of the bond today given that in the next time step there might be a $\delta r$ rise in $r$.
The value of the bond with face value $V_F$ is
\begin{align} \label{Eqn:TBill}
V = \frac{(1-p) V_F}{1+r+\delta r} + \frac{p V_F}{1+r} = (1-p) V_\text{low} + p V_\text{high},
\end{align}
where $p$ and $(1-p)$ denote the probabilities of a constant interest rate and a rise, respectively.
This model is the first step of a binomial tree. 
Binomial trees can be used to price securities with a path dependency such as bonds with embedded options \cite{black1990}.

The simple scenario in Eq. (\ref{Eqn:TBill}) could correspond to a market participant who bought a 1 year T-bill the day before a Federal Open Markets Committee announcement and expects a $\delta r=0.25\%$-points increase of the Federal Funds Rate with a $(1-p) = 70\%$ probability and no change with a $p = 30\%$ probability \footnote{The investor would also have to assume that there is a perfect correlation between one year T-Bills and the Federal Funds Rate. This situation also implies that the market does not expect a change in the Federal Funds Rate to occur.}. 

We show how to calculate the value of the investor's T-bill using the IBM Q Experience by using amplitude estimation and mapping $V$ to $[0, 1]$ such that $V_\text{low}$ and $V_\text{high}$ correspond to $\$0$ and $\$1$, respectively.

Here, we only need a single qubit to represent the uncertainty and the objective and we have $\mathcal{A} = R_y(\theta_p)$, where $\theta_p = 2 \sin^{-1}(\sqrt{p})$, and thus, $\mathcal{A}\ket{0} = \sqrt{1-p}\ket{0} + \sqrt{p}\ket{1}$.

For the one-dimensional case, it can be easily seen that the amplitude estimation operator $Q = \mathcal{A} Z \mathcal{A}^{\dag} Z = R_y(2 \theta_p)$, where $Z$ denotes the corresponding Pauli operator \citep{Nielsen2010}.
We discuss this in more detail in Appendix \ref{sec:q_operator}.
In particular, this implies $Q^{2^j} = R_y(2^{j+1} \theta_p)$, which allows us to construct the amplitude estimation circuit efficiently to approximate the parameter $p = \mathbb{E}[X] = 30\%$.

Although a single period binomial tree is a very simple model, it is straight-forward to extend it to multi-period multi-nomial trees with path-dependent assets.
Thus, it represents the smallest building block for interesting scenarios of arbitrary complexity. 

\subsection*{Results from real quantum hardware}

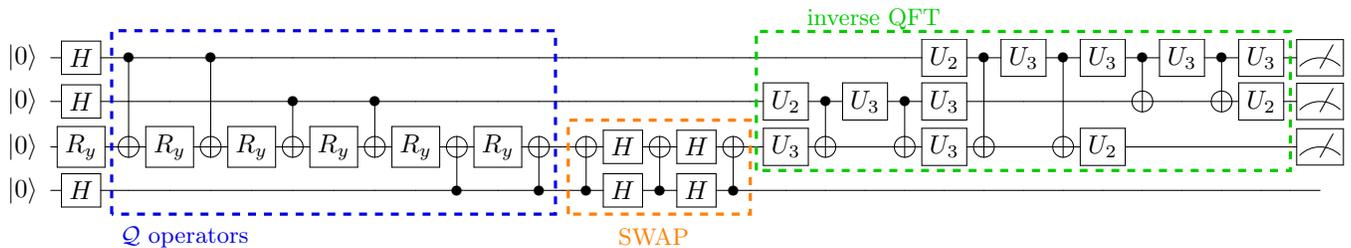
\begin{figure*}[htbp!]
  \begin{tikzpicture}
  \node at (0,0) {\includegraphics[width=\textwidth]{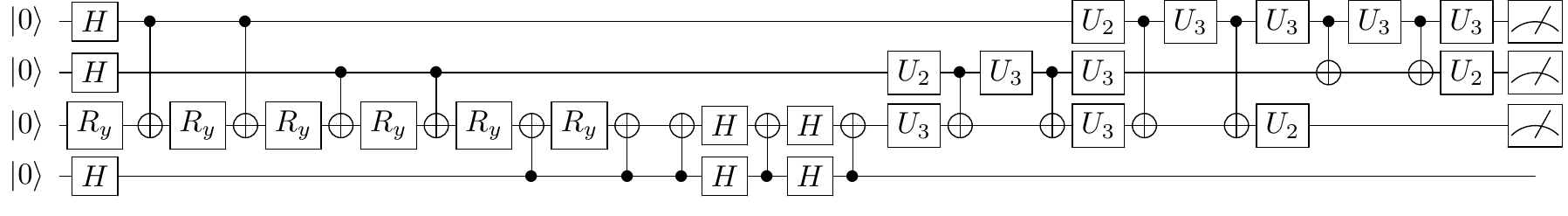}};
  \draw[dashed,blue!90!black,very thick] (-7.47,-1.1) rectangle ++(5.9,2.44);
  \draw[dashed,orange,very thick] (-1.4,-1.1) rectangle ++(2.42,1.25);
  \draw[dashed,green!80!black,very thick] (1.1,-0.52) rectangle ++(7.1,1.85);
  \node[blue!90!black,anchor=west] at (-7.47,-1.4) {$\mathcal{Q}$ operators};
    \node[orange,anchor=west] at (-0.85,-1.4) {SWAP};
  \node[green!80!black,anchor=west] at (1.66,1.5) {inverse QFT};
  \end{tikzpicture}
  \caption{Amplitude estimation circuit for the T-Bill problem with $m=3$. Dashed boxes highlight from left to right: the controlled $Q^{2^j}$, the swap of two qubits, and the inverse QFT. The swap is needed to overcome the limited connectivity of the chip. $U_2$ and $U_3$ indicate single qubit rotations where the parameters are omitted; they are formally introduced in Appendix \ref{sec:u2_u3_rotations}. Note that the circuit could be further optimized, e.g., the adjoint CNOT gates at the beginning of the SWAP would cancel out, but we kept them for illustration.}
  \label{fig:circuit_ae_toy_m3_p30}
\end{figure*}

\begin{figure}[htbp!]
  \includegraphics[width=0.45\textwidth]{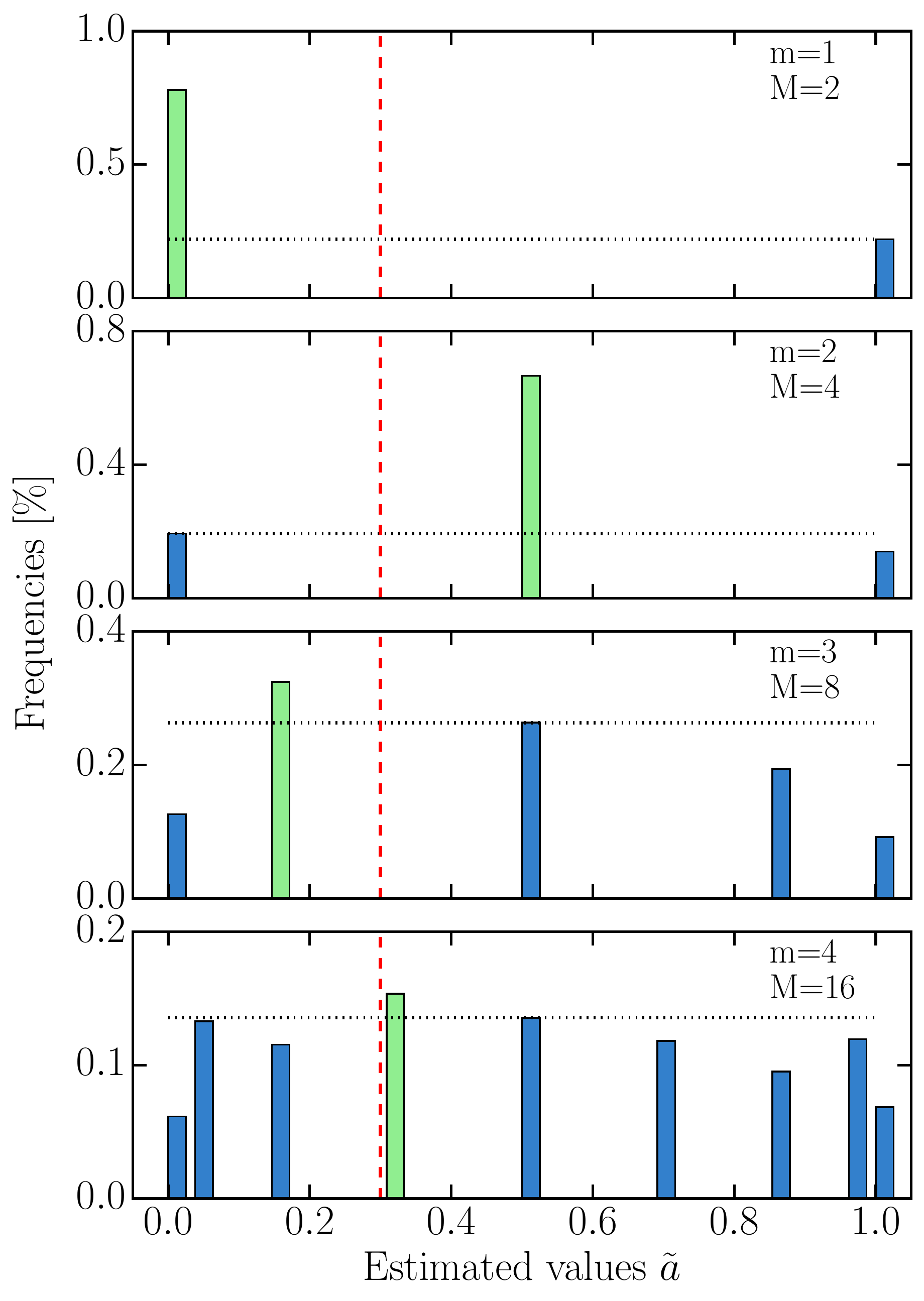}
  \caption{Results of running amplitude estimation on real hardware for $m = 1, ..., 4$ with 8192 shots each. The green bars indicate the probability of the most frequent estimate and the blue bars the probability of the other estimates. The red dashed lines indicate the target value of $30\%$. The gray dashed lines show the probability of the second most frequent value to highlight the resulting contrast. 
The possible values are not equally distributed on the x-axis, since amplitude estimation first returns a number $y \in \{0, ..., M-1\}$ that is then classically mapped to $a = \sin^2\left(\frac{y \pi}{M}\right)$.}
  \label{fig:Toy_Model_p30_Results}
\end{figure}

We run several experiments in which we apply amplitude estimation with a different number of evaluation qubits $m = 1, 2, 3, 4$ corresponding to $M = 2, 4, 8, 16$ samples, respectively, to estimate $p = \mathbb[X]$.
This requires at most five qubits and can be implemented and run on the \emph{IBM Q 5 Yorktown} (ibmqx2) quantum  processor with five qubits accessible via the IBM Q Experience \citep{ibm_q_yorktown_processor}.
As disussed in Sec. \ref{sec:evaluation}, the success probability of amplitude estimation is larger than $8/\pi^2$, but not necessarily $100\%$, and the real hardware introduces additional errors.
Thus, we repeat every circuit $8192$ times (i.e., the maximal number of shots in the IBM Q Experience) to get a reliable estimate.
This implies a constant overhead, which we ignore in the comparison of the algorithms.
The quantum circuit for $m=3$ compiled to the IBM Q 5 quantum processor is illustrated in Fig. \ref{fig:circuit_ae_toy_m3_p30}.
The connectivity of the IBM Q 5 quantum processor, shown in Appendix \ref{sec:ibm_q_topologies}, requires swapping two qubits in the middle of the circuit between the application of the controlled $Q$ operators and the inverse Quantum Fourier Transform.
The results of the algorithm are illustrated in Fig. \ref{fig:Toy_Model_p30_Results} where it can be seen that the most frequent estimator approaches the real value $p$ and how the resolution of the algorithm increases with $m$.
The quantum algorithm presented in this paper outperforms the Monte Carlo method already for $M = 16$ samples (i.e.~$m = 4$ evaluation qubits), which is the largest scenario we performed on the real hardware, see Fig. \ref{fig:convergence_qc_mc}.
The details of this convergence analysis are discussed in Appendix \ref{sec:toy_convergence}.

\begin{figure}[htbp!]
  \includegraphics[width=0.45\textwidth]{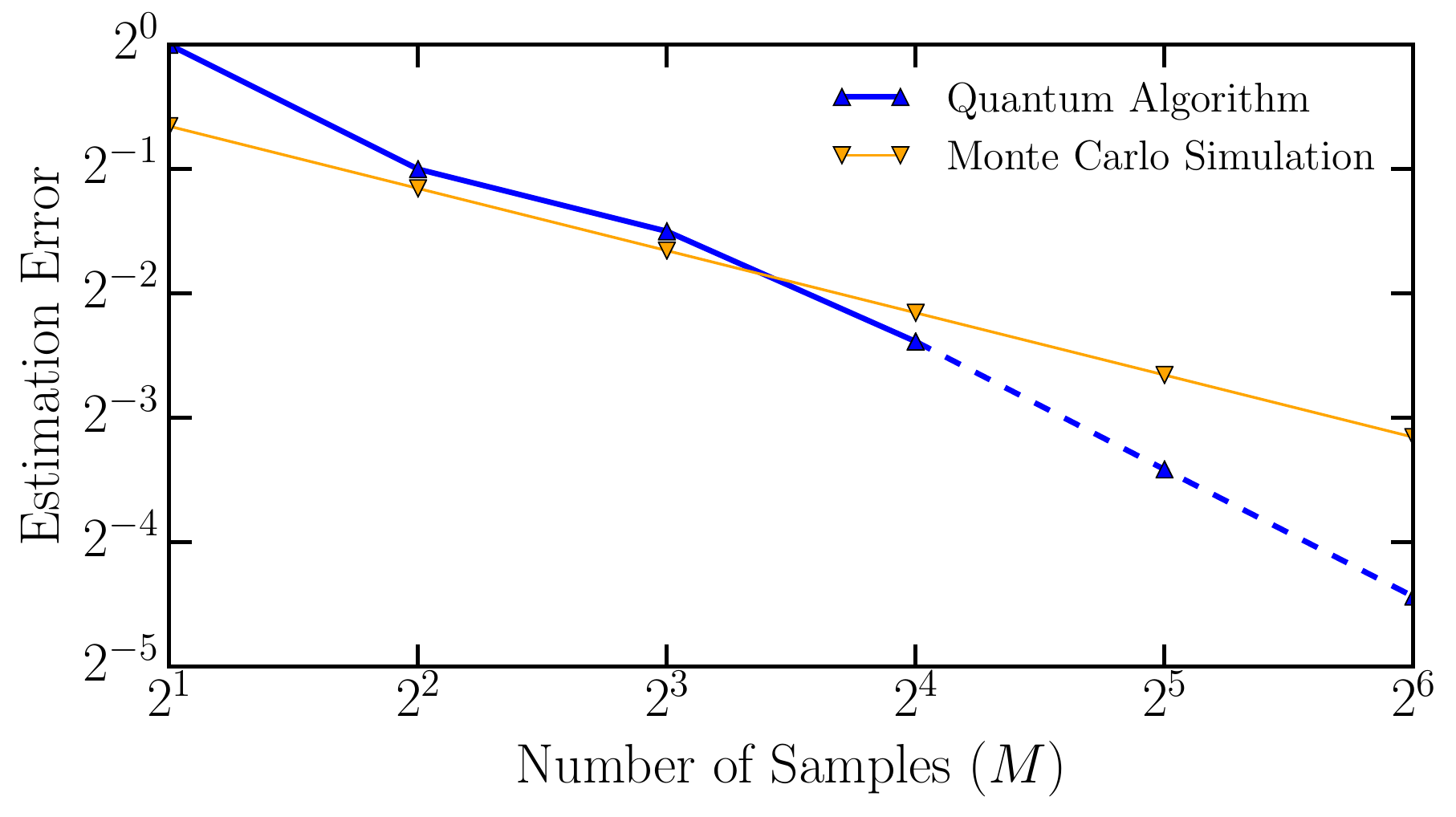}
  \caption{Comparison of the convergence of the error of Monte Carlo simulation and our algorithm with respect to the number of samples $M$. Although the quantum algorithm starts with a larger estimation error, for $M \geq 16$ ($m \geq 4$) the better convergence rate of the quantum algorithm takes over and the error stays below the Monte Carlo results. The green solid line shows the error for our real experiments using up to five qubits. The green dashed line shows how the estimation error would further decrease for experiments with six and seven qubits, respectively.}
  \label{fig:convergence_qc_mc}
\end{figure}


\section{\label{sec:2_asset_problem} Two asset portfolio}

\begin{center}
\begin{figure}[htbp!]
\includegraphics[width=0.49\textwidth]{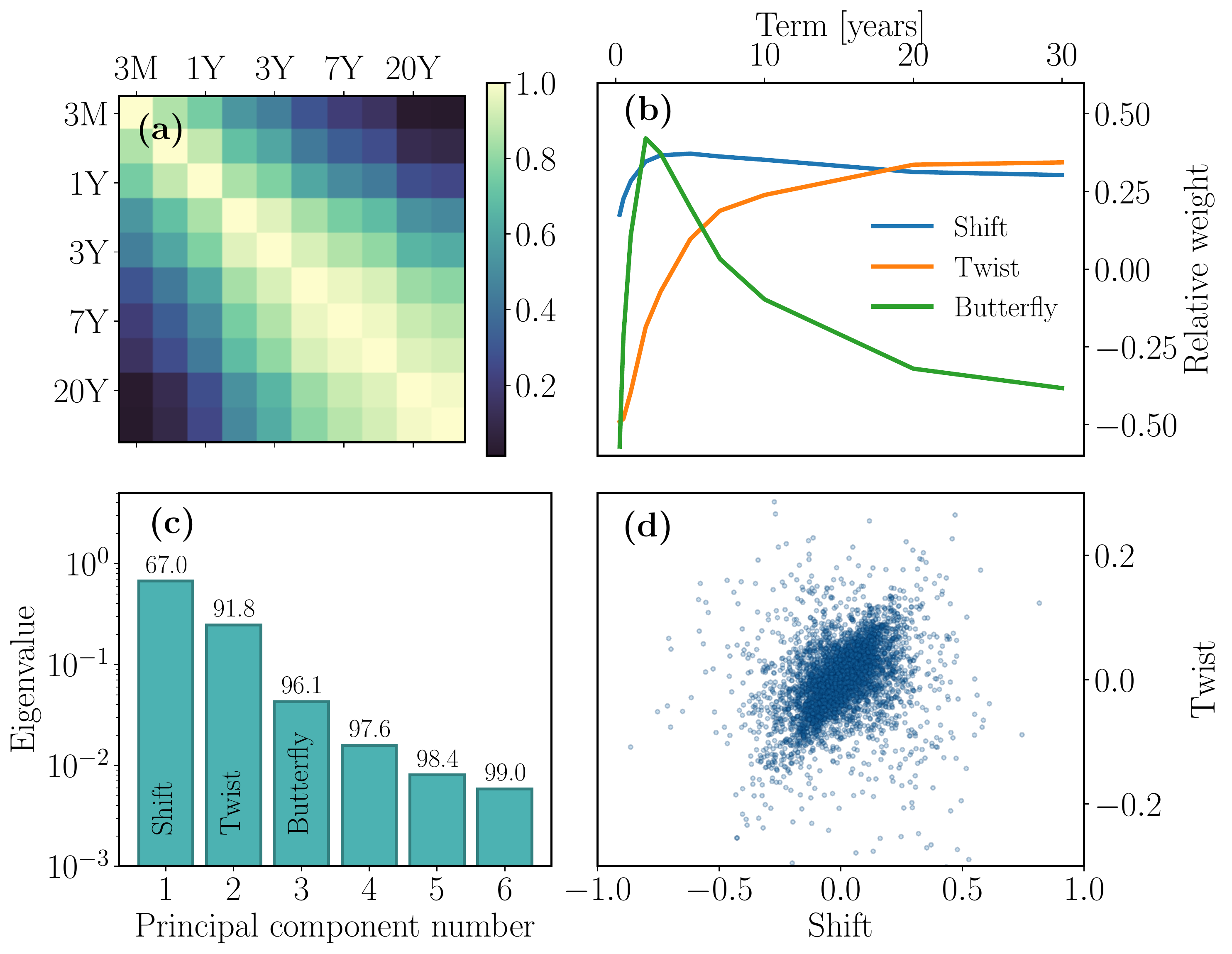}
\caption{\label{Fig:PCA} 
Daily change in the CMT rates. 
(a) Correlation matrix.
The high correlation between the rates can be exploited to reduce the dimension of the problem. 
(b) Shift, Twist and Butterfly components expressed in terms of the original constant maturity treasury rates.
(c) Eigenvalues of the principal components. The numbers show the cumulative explained variance.
(d) Marginal distribution of the Shift vs the Twist principal components.
}
\end{figure}
\end{center}

We now illustrate how to use our algorithm to calculate the daily risk in a portfolio made up of one-year US Treasury bills and two-year US Treasury notes with face values $V_{F_1}$ and $V_{F_2}$, respectively. 
We chose a simple portfolio in order to put the focus on the amplitude estimation algorithm applied to VaR. 
The portfolio is worth
\begin{align}
V(r_1,r_2)=\frac{V_{F_1}}{1+r_1}+\sum_{i=1}^{4}\frac{c V_{F_2}}{(1+r_2/2)^i}+\frac{V_{F_2}}{(1+r_2/2)^4} \label{eq:portfolio_value}
\end{align}
where $c$ is the coupon rate paid every six months by the two-year treasury note and $r_1$ and $r_2$ are the yield to maturity of the one-year bill and two-year note, respectively.
US Treasuries are usually assumed to be default free \cite{Nippani2001}. 
The cash-flows are thus known \emph{ex ante} and the changes in the interest rates are the primary risk factors.
Therefore, a proper understanding of the yield curve suffices to model the risk in this portfolio. 
In this work we use the Constant Maturity Treasury (CMT) rates to model the uncertainty in $r_1$ and $r_2$, see Appendix \ref{sec:data} for a description of the data. 
To calculate the daily risk of our portfolio we study the difference in the CMT rates from one day to the next. 
These differences are highly correlated (as are the initial CMT rates), see Fig. \ref{Fig:PCA}(a), making it unnecessary to model them all when simulating more complex portfolios. 
A principal component analysis reveals that the first three principal components, named shift, twist and butterfly account for 96\% of the variance \cite{Colin2006, Vannerem2010}, see Fig. \ref{Fig:PCA}(b)-(d). 
Therefore, when modeling a portfolio of US Treasury securities it suffices to study the distribution of these three factors. 
This dimensionality reduction also lowers the amount of resources needed by our quantum algorithm.

\begin{figure}[ht]
	\centering
		\includegraphics[width=.25\textwidth]{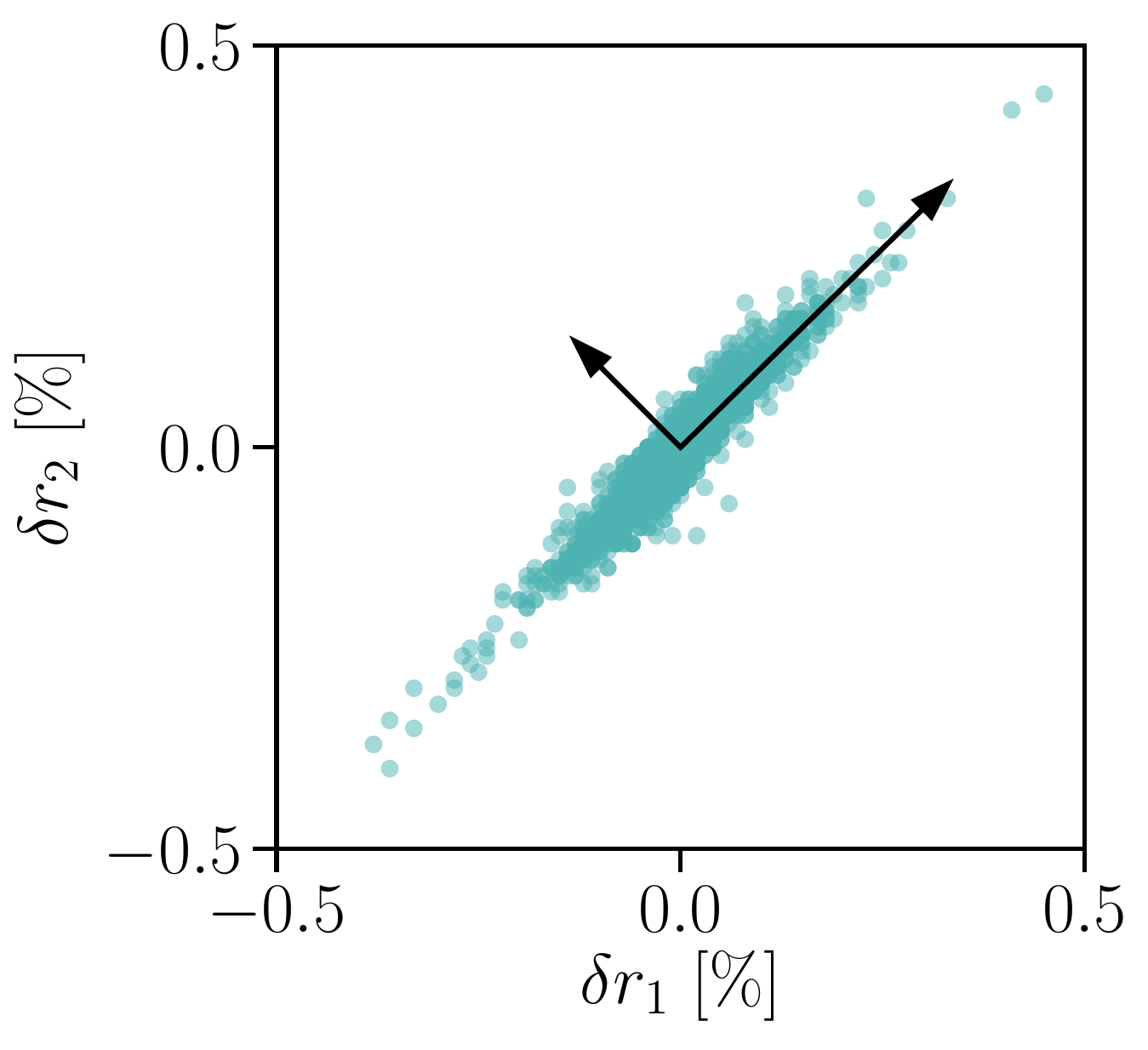}
	\caption{Historical constant maturity treasury rates (1-year against 2-years to maturity) as well as the resulting principal components: shift (longer vector), and twist (shorter vector).}
	\label{fig:Y1_Y2_PCA}
\end{figure}

To study the daily risk in the portfolio we write $r_i=r_{i,0}+\delta r_i$ where $r_{i,0}$ is the yield to maturity observed today and the random variable $\delta r_i$ follows the historical distribution of the one day changes in the CMT rate with maturity $i$.
For our demonstration we set $V_{F_1} = V_{F_2} = \$100$, $r_{1,0} = 1.8\%$, $r_{2,0} = 2.25\%$, and $c = 2.5\%$ in Eq. (\ref{eq:portfolio_value}).
We perform a principal component analysis of $\delta r_1$ and $\delta r_2$ and retain only the shift $S$ and twist $T$ components. 
Figure \ref{fig:Y1_Y2_PCA} illustrates the historical data as well as $S$ and $T$, related to $\delta r_i$ by
\begin{align}
\begin{pmatrix} \delta r_1 \\ \delta r_2 \end{pmatrix} =
\boldsymbol{W} \begin{pmatrix} S \\ T \end{pmatrix} =
\begin{pmatrix} 0.703 & -0.711 \\ 0.711 & 0.703 \end{pmatrix} \begin{pmatrix} S \\ T \end{pmatrix}.
\end{align}
The correlation coefficient between shift and twist is $-1\%$.
We thus assume them to be independent and fit discrete distributions to each separately, see Fig. \ref{fig:Distribution_Fitting}.
We retained only the first two principal components to illustrate the use of principal component analysis despite the fact that, in this example, there is no dimensionality reduction.
Furthermore, this allows us to simulate our algorithm in a reasonable time on classical hardware by keeping the number of required qubits low.
We expect that all three components would be retained when running this algorithm on real quantum hardware for larger portfolios.

\subsection{Uncertainty representation in the quantum computer \label{sec:two_asset_uncertainty}}

\begin{figure}[ht]
	\centering
		\includegraphics[width=.48\textwidth]{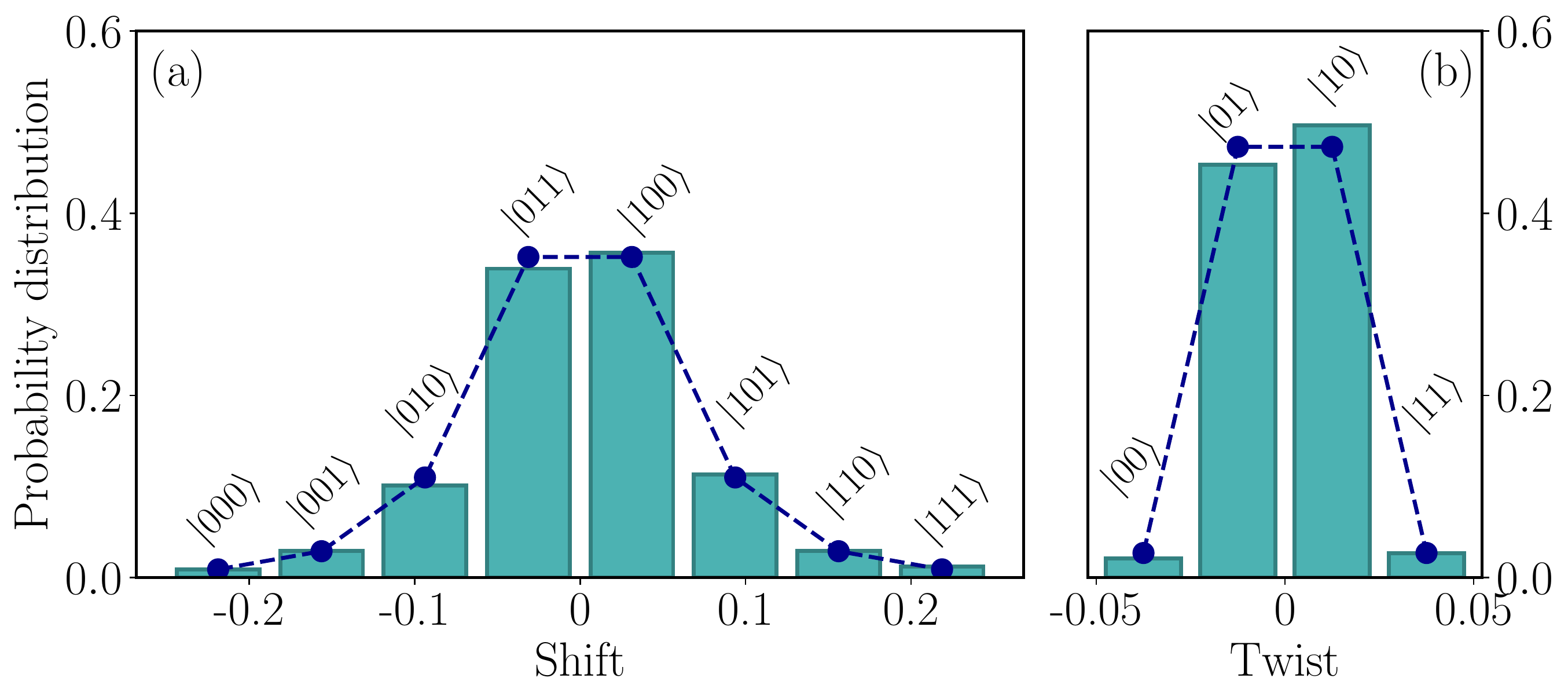}
		\includegraphics[width=.48\textwidth]{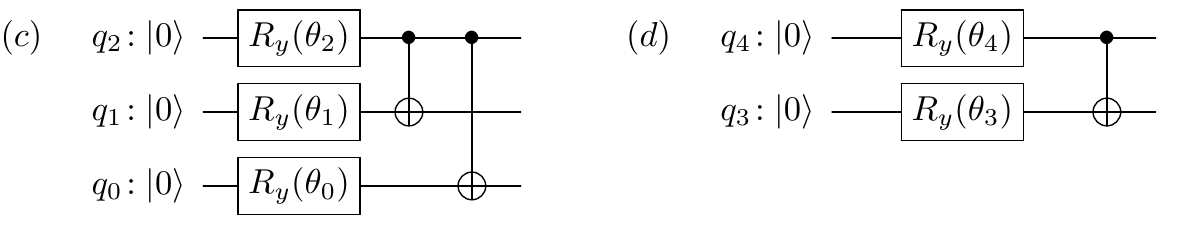}
	\caption{
	(a) 8-bin histogram of historical shift data (bars) as well as fitted distribution (dashed line). 
	(b) 4-bin histogram of historical twist data (bars) as well as fitted distribution (dashed line). In both cases the labels show the quantum state that will occur with the corresponding probability. (c) and (d) show the quantum circuits used to load the distributions of (a) and (b), respectively, into the quantum computer.}
	\label{fig:Distribution_Fitting}
\end{figure}

We use three qubits, denoted by $q_0, q_1, q_2$, to represent the distribution of $S$,  and two, denoted by $q_3, q_4$, for $T$. 
As discussed in Sec. \ref{sec:mapping}, the probability distributions are encoded by the states $\ket{\psi_S}=\sum_{i=0}^{7}\sqrt{p_{i,S}}\ket{i}_8$ and $\ket{\psi_T}=\sum_{i=0}^{3}\sqrt{p_{i,T}}\ket{i}_4$ for $S$ and $T$, which can thus take eight and four different values, respectively.
We use more qubits for $S$ than for $T$ since the shift explains a larger part of the variance. Additional qubits may be used to represent the probability distributions at a higher resolution.
The qubits naturally represent integers via binary encoding and we apply the affine mappings
\begin{eqnarray}
S &=& 0.0626\, x - 0.2188, \\
T &=& 0.0250\, y - 0.0375. \label{eq:affine}
\end{eqnarray}
Here $x \in \{0, ..., 7\}$ and $y \in \{0, ..., 3\}$ denote the integer representations of $S$ and $T$, respectively.
Given the almost perfect symmetry of the historical data we fit symmetric distributions to it. 
The operator $\mathcal{R}$ that we define prepares a quantum state $\mathcal{R}\ket{0}_5$, illustrated by the dots in Fig. \ref{fig:Distribution_Fitting}, that represents the distributions of $S$ and $T$, up to the aforementioned affine mapping.



\subsection{Portfolio model on the quantum computer \label{sec:two_portfolio_model}}

Next, we show how to construct the operator $F$ to translate the random variables $x$ and $y$ into a portfolio value. Equations (\ref{eq:portfolio_value}) through (\ref{eq:affine}) allow us to define the portfolio value $V$ in terms of $x$ and $y$, instead of $r_1$ and $r_2$.
For simplicity, we use a first order approximation
\begin{align} \label{Eq:fprime}
\tilde{f}(x, y) = 203.5170 - 13.1896 x - 1.8175 y
\end{align}
of $V$ around the mid points $x = 3.5$ and $y = 1.5$.
From a financial perspective, the first order approximation $\tilde{f}$ of $V$ corresponds to studying the portfolio from the point of view of its duration \cite{Martellini2003}. 
Higher order expansions, e.g. convexity could be considered at the cost of increased circuit depth. 

To map the approximated value of the portfolio $\tilde{f}$ to a function $f$ with target set $[0, 1]$ we compute $f = (\tilde{f} - \tilde{f}_{\min})/(\tilde{f}_{\max}-\tilde{f}_{\min})$, where $\tilde{f}_{\min} = \tilde{f}(7, 3)$ and $\tilde{f}_{\max} = \tilde{f}(0, 0)$, i.e., the minimum and maximum values $\tilde{f}$ can take for the considered values of $x \in \{0, ..., 7\}$ and $y \in \{0, ..., 3\}$.
This leads to
\begin{align} \label{Eq:fprime}
f(x, y) = 1 - 0.1349 x - 0.0186 y.
\end{align}
The approach, illustrated in Fig. \ref{fig:polynomial_circuit}, allows us to construct an operator $F$ corresponding to $f$ for a given scaling parameter $c \in (0, 1]$.

\subsection{Results from simulations of an ideal quantum computer}

We simulate the two-asset portfolio for different numbers $m$ of sampling qubits to show the behavior of the accuracy and convergence rate. 
We repeat this task twice, once for a processor with all-to-all connectivity and once for a processor with a connectivity corresponding to the IBM Q 20 chip, see Appendix \ref{sec:ibm_q_topologies}.
This highlights the overhead imposed by a realistic chip connectivity.
For a number $M=2^m$ samples, we need a total of $m+12$ qubits for expected value and VaR, and $m+13$ qubits for CVaR. Five of these qubits are used to represent the distribution of the interest rate changes, see Sec. \ref{sec:two_asset_uncertainty}, one qubit is needed to create the state in Eq. (\ref{eq:f_operator}) used by amplitude estimation, and six ancillas are needed to implement the controlled $Q$ operator. For CVaR we need one more ancilla for the comparison to the level $l$ as discussed in Sec. \ref{sec:mapping}.
Once the shift and twist distributions are loaded into the quantum computer, using the circuit shown in Fig. \ref{fig:Distribution_Fitting}(c) and (d), we apply the operator $F$ to create the state defined in Eq. (\ref{eq:f_operator}). 

\begin{figure}[htbp!]
  \includegraphics[width=0.45\textwidth]{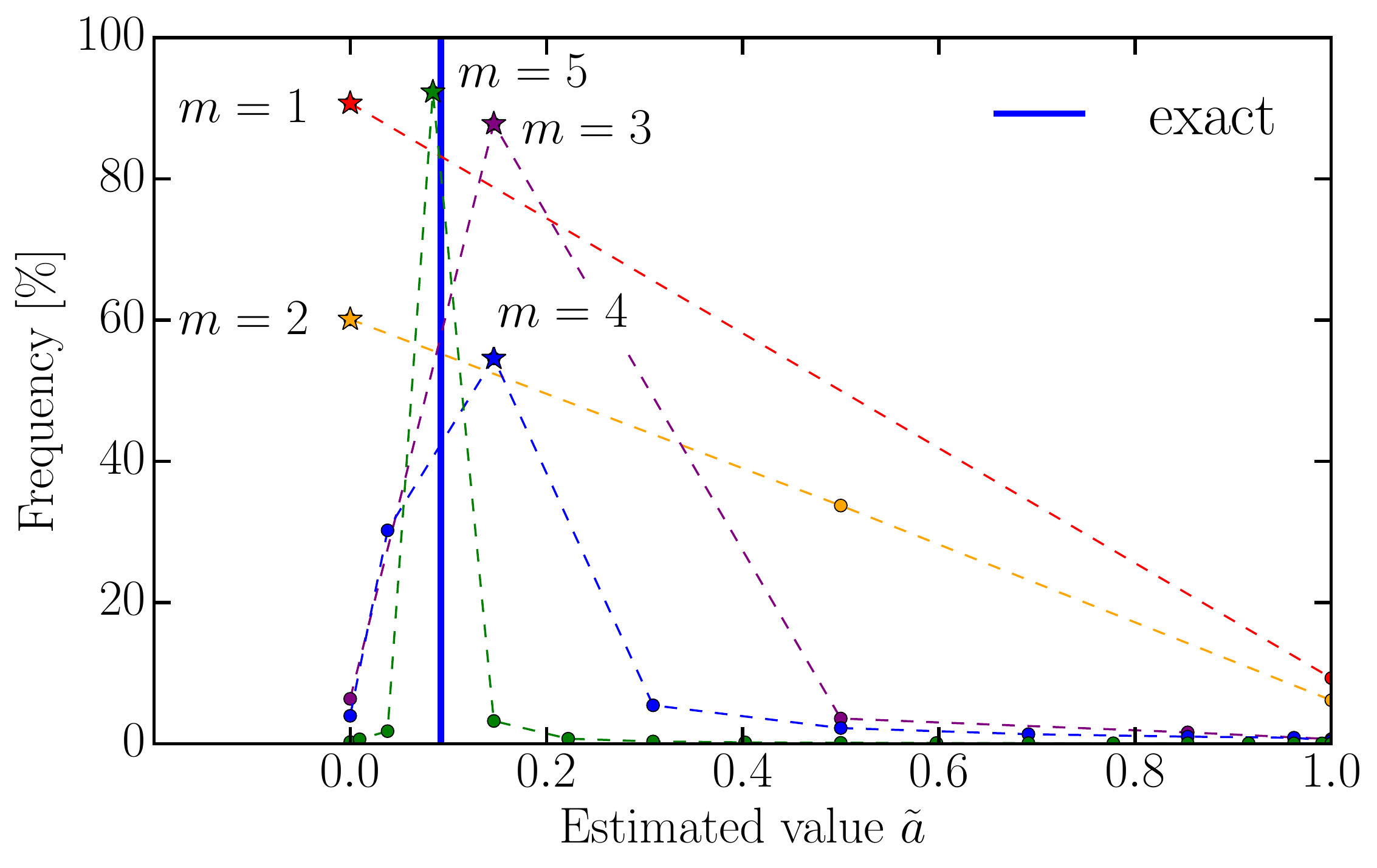}
  \caption{VaR estimated through a simulation of a perfect quantum computer. As the number of sample qubits $m$ is increased the quantum estimated VaR approaches the classical value indicated by the vertical blue line. The dashed lines are intended as guides to the eye. The stars indicate the most probable values.}
  \label{fig:cvar_perfect_results}
\end{figure}

We compare the quantum estimation of risk to the exact $95\%$ VaR level of $\$0.288$.
When taking into account the mapping of Sec. \ref{sec:two_portfolio_model}, this classical VaR corresponds to 0.093, shown by the verticle line in Fig. \ref{fig:cvar_perfect_results}.
The quantum estimation of risk rapidly approaches this value as $m$ is increased, see Fig. \ref{fig:cvar_perfect_results}. 
With $m=5$ sample qubits the difference between the classical and quantum estimates is 9\%.
The number of CNOT gates needed to calculate VaR approximately doubles each time a sample qubit is added, see Tab. \ref{tab:gate_counts}, i.e. it scales as $O(M)$ with a resulting error of $O(M^{-1})$.

We find that the connectivity of the IBM Q 20 chip increases  the number of CNOT gates by a factor $2.5$ when compared to a chip with all-to-all connectivity
\footnote{These results are based on QISKit 0.5, future version might be able to further reduce the CNOT overhead, cf.~\cite{Qiskit}}.

\begin{table}
\begin{center}
\caption{\label{tab:gate_counts}Summary of the number of CNOT gates to estimate VaR as a function of $m$ for a processor architecture featuring an all-to-all qubit connectivity and an architecture with a qubit connectivity corresponding to the IBM Q 20 Austin chip with 20 qubits.}
\begin{tabular}{c c r r r r r r c} \hline\hline
  &    &          & \multicolumn{3}{c}{\#CX} \\
m &  M & \#qubits & \hspace{0.5cm} all-to-all & \hspace{0.5cm} IBM Q 20 & \hspace{0.5cm} overhead \\ \hline
1 &  2 &       13 &    795 &   1'817 & 2.29 \\
2 &  4 &       14 &  2'225 &   5'542 & 2.49 \\
3 &  8 &       15 &  5'085 &  12'691 & 2.50 \\
4 & 16 &       16 & 10'803 &  26'457 & 2.45\\
5 & 32 &       17 & 22'235 &  55'520 & 2.50 \\ \hline\hline
\end{tabular}
\end{center}
\end{table}

\subsection{Results from simulations of a noisy quantum computer}

Computing risk for the two-asset portfolio requires a long circuit.
However, it suffices for amplitude estimation to return the correct state with the highest probability, i.e. measurements do not need to yield this state with 100\% probability. 
We now run simulations with errors to investigate how much imperfections can be tolerated before the correct state can no longer be identified.

We study the effect of two types of errors: energy relaxation and cross-talk, where the latter is only considered for two-qubit gates (CNOT gates). 
We believe this to be a sufficient approximation to capture the leading error sources. Errors and gate times for single qubit gates are in general an order of magnitude lower than for two-qubit gates \cite{Sheldon2016b, Gustavsson2013, QuantumExperience}. Furthermore, our algorithm requires the same order of magnitude in the number of single and two-qubit gates, see Tab. \ref{tab:gate_counts}.
Energy relaxation is simulated using a relaxation rate $\gamma$ such that after a time $t$ each qubit has a probability $1-\exp(-\gamma t)$ of relaxing to $\ket{0}$ \cite{Qiskit}.
We set the duration of the CNOT gates to $100~\rm{ns}$ and assume that the single qubit gates are done instantly and are thus exempt from errors.
We also include qubit-qubit cross-talk in our simulation by adding a $ZZ$ error-term in the generator of the CNOT gate
\begin{align}
\exp\{-i\pi(ZX+\alpha ZZ)/4\}.
\end{align}
Typical cross-resonance \cite{Rigetti2010} CNOT gate rates are of the order of $5~\rm{MHz}$ whilst cross-talk on IBM Q chips are of the order of $-100~\rm{kHz}$ \cite{QuantumExperience}. 
We thus estimate a reasonable value of $\alpha$, i.e. the strength of the cross-talk, to be $-2\%$ and simulate its effect over the range $[-3\%,\,0\%]$.

\begin{figure}[ht]
	\centering
		\includegraphics[width=.45\textwidth]{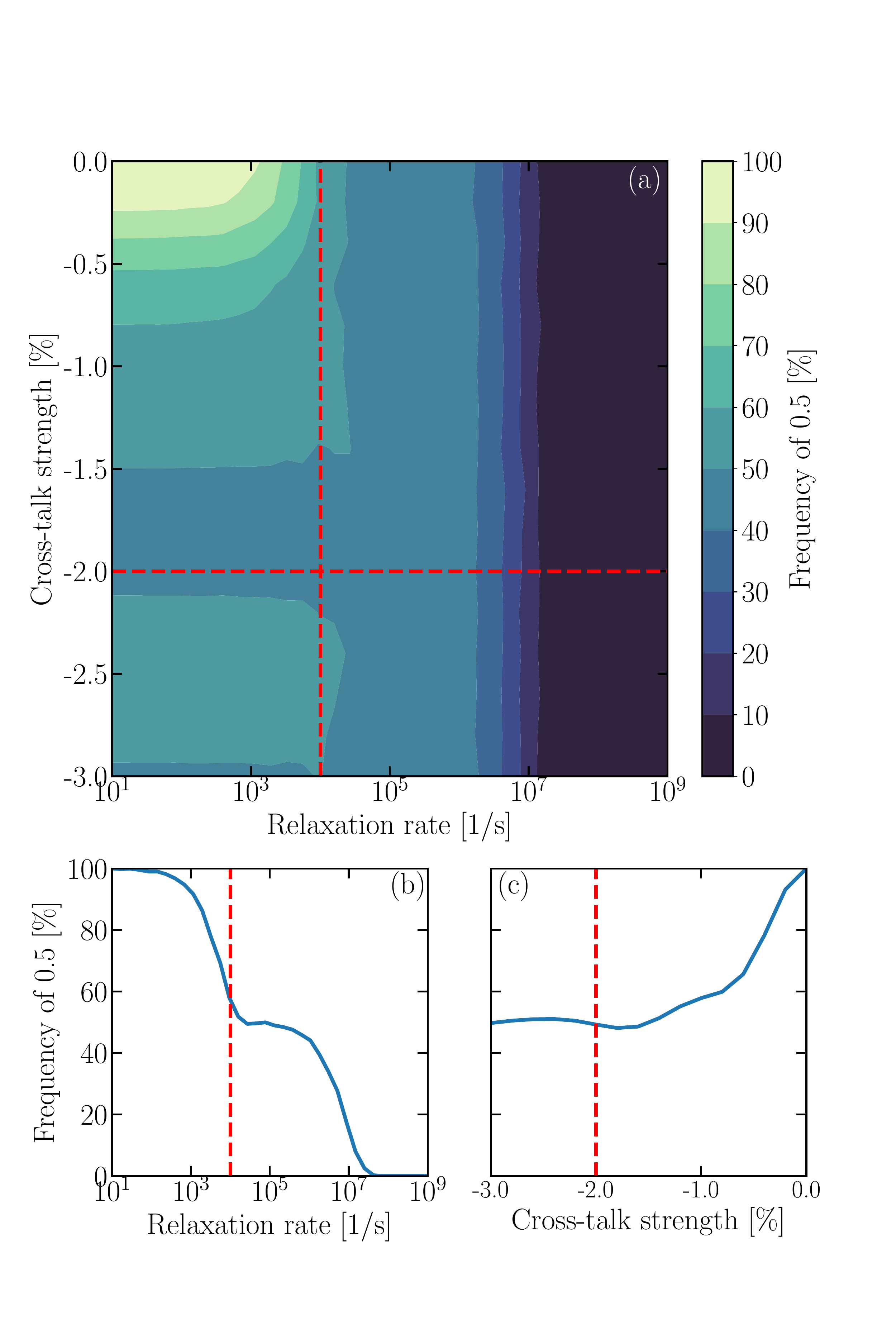}
	\caption{Results from noisy simulation for estimating the expected value of the two-asset portfolio using two evaluation qubits. The perfect simulation returns $0.5$ with 100$\%$. This figure shows how the probability of measuring $0.5$ decreases with increasing noise: (a) shows the results for both, increasing cross-talk and increasing relaxation rate, (b) shows the result for varying relaxation rate without cross-talk, and (c) shows the result for different cross-talk strengths without relaxation. The dashed red lines indicate the estimated state of the currently available hardware.}
	\label{fig:noisy_results}
\end{figure}

We illustrate the effect of these errors by computing the expected value of the portfolio.
Since the distributions are symmetric around zero and mapped to the interval $[0,1]$ we expect a value of $0.5$, i.e. from one day to the next we do not expect a change in the portfolio value.
This simulation is run with $m=2$ sample qubits since this suffices to exactly estimate $0.5$.
The algorithm is successful if it manages to identify 0.5 with a probability greater than 50\%.
With our error model this is achieved for relaxation rates $r\gamma < 10^{-4}~\rm{s}^{-1}$ and cross-talk strength $\vert\alpha\vert<1\%$, see Fig. \ref{fig:noisy_results}(a)-(c), despite the 4383 gates needed.
A generous estimation of current hardware capabilities with $\gamma=10^{-4}~\rm{s}^{-1}$ (loosely based on $T_1=100~\mu\rm{s}$) and $\alpha=-2\%$, shown as red lines in Fig. \ref{fig:noisy_results}, indicates that this simulation may be possible in the near future as long as other error sources (such as measurement error and unitary errors resulting from improper gate calibrations) are kept under control.

\section{\label{sec:conclusion} Conclusion}

We developed a quantum algorithm to estimate risk, e.g. for portfolios of financial assets, resulting in a quadratic speedup compared to classical Monte Carlo methods.
The algorithm has been demonstrated on real hardware for a small model and the scalability and impact of noise has been studied using a more complex model and simulation.
Our approach is very flexible and straight-forward to extend to other risk measures such as semi-variance.

More qubits are needed to model realistic scenarios and the errors of actual hardware need to be reduced.
Although the quadratic speedup can already be observed for a small number of samples, more is needed to achieve a practical quantum advantage.
In practice, Monte Carlo simulations can be massively parallelized, which pushes the border for a quantum advantage even higher.

Our simulations of the two-asset portfolio show that circuit depth is limited for current hardware.
In order to perform the calculation of VaR for the two asset portfolio on real quantum hardware it is likely that qubit coherence times will have to be increased by several orders of magnitude and that cross-talk will have to be further suppressed.

However, approximating, parallelizing, and decomposing quantum phase estimation is ongoing research and we expect significant improvements in this area not only through hardware, but also algorithms \citep{Dobsicek2006, OLoan2010, Svore2013}.
This can also help to shorten the required circuit depths, and thus, to reduce the requirements on the hardware to achieve a quantum advantage.
Circuit depth can also be shortened by using a more versatile set of gates.
For instance, the ability to implement SWAP gates directly in hardware would circumvent the need to synthesize them using CNOT gates \cite{Egger2018b, Sjoqvist2012}.
In addition, techniques such as error mitigation \cite{Kandala2018} could be applied to cope with the noisy hardware of the near future.

Another question that has only briefly been addressed in this paper is the loading of considered random distributions or stochastic processes.
For auto-correlated processes this can be rather costly and needs to be further investigated.
Techniques known from classical Monte Carlo, such as importance sampling \citep{Tokdar2010}, might be employed here as well to improve the results or reduce the circuit depth.

\begin{acknowledgments}
We want to thank Lior Horesh for his insights and the stimulating discussions. IBM and IBM Q are trademarks of International Business Machines Corporation, registered in many jurisdictions worldwide. Other product or service names may be trademarks or service marks of IBM or other companies.
\end{acknowledgments}

\appendix

\section{\label{sec:q_operator} $Q$-Operator}

For a given circuit $\mathcal{A}$ acting on $n+1$ qubits, the corresponding $Q$-operator used in amplitude estimation is defined as \citep{Brassard2000}
\begin{eqnarray*}
Q 
&=& \mathcal{A}(\mathbb{I} - 2 \ket{0}_{n+1} \bra{0}_{n+1})\mathcal{A}^{\dagger}  \\
&& (\mathbb{I} - 2 \ket{\psi_0}_{n}\ket{0} \bra{\psi_0}_{n}\bra{0}),
\end{eqnarray*}
where $\mathbb{I}$ denotes the identity operator.
If $n=0$, as e.g.~considered in Sec. \ref{sec:1_asset_problem}, the reflections defining $Q$ reduce to the Pauli $Z$-operators and $Q$ simplifies to $\mathcal{A}Z\mathcal{A}^{\dagger}Z$. In addition, if $\mathcal{A} = R_y(\theta)$ then it can be easily seen that $Q = R_y(2\theta)$.

\section{\label{sec:taylor_error_bound} Error bound for $F$ approximation}

Suppose $p(x)$ denotes the Taylor approximation of Eq. (\ref{eq:polynomial_definition}) of order $(2u+1)$.
Then, the error bound is derived from the next coefficient in the Taylor series (plus higher order terms).
Therefore, we analyze the Taylor series
\begin{eqnarray*}
&& \sin^{-1}\left(\sqrt{y + \frac{1}{2}}\right) \\ 
&=& \frac{\pi}{4} + \sum_{u=0}^\infty \left( \prod_{i=1}^u (2 i - 1) \right) \frac{2^u}{(2u+1)u!} y^{2u+1},
\end{eqnarray*}
for $y \in [-\frac{1}{2}, +\frac{1}{2}]$.
The Taylor series can be derived by first taking the derivative, using the corresponding Taylor series, and integrating the different terms independently.
Replacing $y$ by $c y$ for $c \in (0, 1]$ and the fact that the extreme values are assumed for $y=\pm \frac{1}{2}$ leads to a bound on the individual terms for a particular $u$ given by
\begin{eqnarray*}
\left( \prod_{i=1}^u (2 i - 1) \right) \frac{2^u}{(2u+1) 2^{2u+1} u!} c^{2u+1} 
& \leq &
\frac{c^{2u+1}}{(2u+1) 2^{u}},
\end{eqnarray*}
where we used $\prod_{i=1}^u (2 i - 1) \leq 2u!$.
For a Taylor approximation of order $2u+1$, the error bound is given by the bound on the next Taylor coefficient in the series, i.e. for $2u+3$.

\section{\label{sec:cvar_error_bound} CVaR Error Bound}

Since $\mathbb{P}[X \leq l_{\alpha}]$ is replaced by an estimation, we cannot directly apply the amplitude estimation error bound for CVaR.
Assume two unknowns $A, B > 0$ and  their estimates $\tilde{A} = A + \delta_a$, $\tilde{B} = B + \delta_b > 0$, where $|\delta_a|, |\delta_b| \leq \delta$ for $\delta > 0$. 
The first order Taylor approximation of $\frac{A}{B} - \frac{\tilde{A}}{\tilde{B}}$ with respect to $\delta_a$ and $\delta_b$ around zero can be used to derive
\begin{eqnarray}
\left|\frac{A}{B} - \frac{\tilde{A}}{\tilde{B}}\right| 
\leq \frac{1}{B}\left|\delta_a - \frac{A}{B} \delta_b\right|
\leq \frac{1}{B} \left(1 + \frac{A}{B} \right) \delta,
\end{eqnarray}
where we ignore higher order terms of $\delta_a$ and $\delta$.

Setting $\frac{A}{B} = \frac{1}{l_{\alpha}} \text{CVaR}_{\alpha}(X)$ and $B = \mathbb{P}[X \leq l_{\alpha}]$, multiplying everything with $l_{\alpha}$ and replacing $\delta$ by $\frac{\pi}{M}$ leads to the following bound for the approximation error $\epsilon > 0$ of $\text{CVaR}_{\alpha}(X)$:
\begin{eqnarray}
\epsilon & \leq & \frac{ l_{\alpha} + \text{CVaR}_{\alpha}(X) }{\mathbb{P}[X \leq l_{\alpha}]} \frac{\pi}{M} \\
&\approx & \frac{ \text{VaR}_{\alpha}(X) + \text{CVaR}_{\alpha}(X) }{1 - \alpha} \frac{\pi}{M},
\end{eqnarray}
where again we omit higher order terms.
Thus, the quantum estimation of CVaR also achieves a quadratic speedup compared to classical Monte Carlo methods.

\section{\label{sec:u2_u3_rotations} $U_2$, $U_3$ single qubit rotations}

In the following, we define the single qubit rotations, $U_2$, $U_3$, used in Fig. \ref{fig:circuit_ae_toy_m3_p30}, and defined e.g. in \citep{Qiskit}:
\begin{eqnarray*}
U_2(\phi, \lambda) &=& 
\left(\begin{array}{cc} 
1/\sqrt{2} & -e^{i \lambda}/\sqrt{2} \\ 
e^{i \phi}/\sqrt{2} & e^{i \lambda + i \phi}/\sqrt{2}
\end{array}\right) \\
U_3(\theta, \phi, \lambda) &=&
\left(\begin{array}{cc} 
\cos(\theta/2) & -e^{i \lambda} \sin(\theta/2) \\
e^{i \phi} \sin(\theta/2) & e^{i \lambda + i \phi} \cos(\theta/2)
\end{array}\right).
\end{eqnarray*}

\section{\label{sec:toy_convergence} Convergence Analysis}

For the Monte Carlo simulation we consider a $95\%$-confidence interval.
To enable a fair comparison despite the small number of samples, we compute an optimistic bound and assume the exact standard error $\sqrt{p (1 - p)}$, where $p$ denotes the success probability of the Bernoulli random variable.
For $p = 0.3$ and a $95\%$ confidence level, the resulting confidence interval is given by $[0.3 - 0.898 / \sqrt{M}, 0.3 + 0.898 / \sqrt{M}]$.

For the quantum algorithm we exploit the error bound given in Eq. (\ref{eq:approx_error_1}).
Although the exact value of $p$ is supposed to be unknown, we can use the estimated value $\tilde{a}$ to compute an error bound.
The algorithm results in an integer $y$ which is classically mapped to $\tilde{a} = \sin^2(y\pi/M)$, and we assume $y$ is the most probable result of the quantum algorithm.
The theory says that for $\theta_a$, defined through $a = \sin^2(\theta_a)$, it holds that $\theta_a \in [(y-1)\pi/M, (y+1)\pi/M]$.
Then, the interval for $\theta_a$ can be mapped to an interval for $a$, whose width is compared to the confidence interval from the Monte Carlo simulation.
Since the mapping from $y$ to $a$ is non-linear, the error bound is not symmetric around $\tilde{a}$ and we consider the maximum.
\newline
\section{\label{sec:data} Data used in this work}

In this work we use the Constant Maturity Treasury rates obtained from the U.S. Department of the Treasury \cite{treasury}. The data is made up of 1/4, 1/2, 1, 2, 3, 5, 7, 10, 20 and 30 year rates resulting from an interpolation of the daily yield curve obtained from the bid yield of actively traded treasury securities at market close.
We only consider periods where all rates are available and ignore the others.
In total, we use more than $5'000$ data points.

\section{\label{sec:ibm_q_topologies} Topology of IBM Q 5 and IBM Q 20 quantum processor}

\begin{figure}[ht]
	\centering
		\includegraphics[width=.2\textwidth]{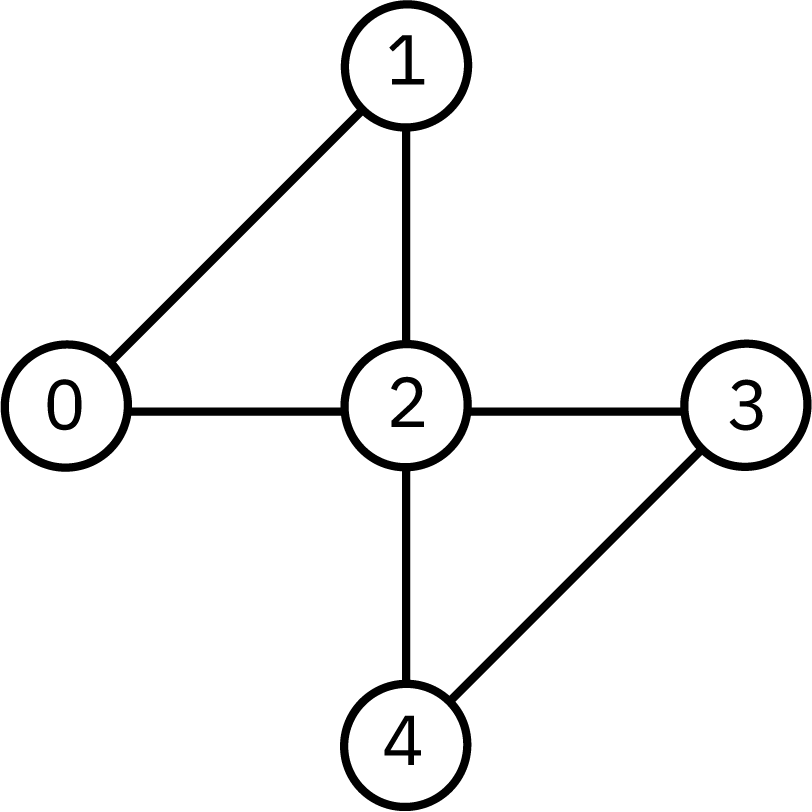}
	\caption{Topology of IBM Q 5 quantum processor.}
	\label{fig:ibm_q_5_topology}
\end{figure}

\begin{figure}[ht]
	\centering
		\includegraphics[width=.3\textwidth]{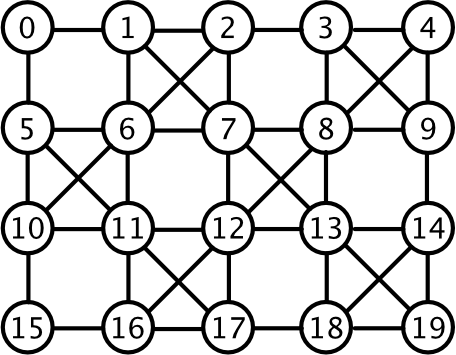}
	\caption{Topology of IBM Q 20 quantum processor.}
	\label{fig:ibm_q_20_topology}
\end{figure}

Figures \ref{fig:ibm_q_5_topology} and \ref{fig:ibm_q_20_topology} show the topologies of the IBM Q 5 \citep{ibm_q_yorktown_processor} and IBM Q 20 \citep{ibm_q_austin_processor} quantum processors.
The lines indicate the connectivity, i.e., the pairs of qubits that allow the application of CNOT gates.

\bibliography{quantum_risk_analysis}

\begin{thebibliography}{49}
\expandafter\ifx\csname natexlab\endcsname\relax\def\natexlab#1{#1}\fi
\expandafter\ifx\csname bibnamefont\endcsname\relax
  \def\bibnamefont#1{#1}\fi
\expandafter\ifx\csname bibfnamefont\endcsname\relax
  \def\bibfnamefont#1{#1}\fi
\expandafter\ifx\csname citenamefont\endcsname\relax
  \def\citenamefont#1{#1}\fi
\expandafter\ifx\csname url\endcsname\relax
  \def\url#1{\texttt{#1}}\fi
\expandafter\ifx\csname urlprefix\endcsname\relax\def\urlprefix{URL }\fi
\providecommand{\bibinfo}[2]{#2}
\providecommand{\eprint}[2][]{\url{#2}}

\bibitem[{\citenamefont{Glasserman et~al.}(2000)\citenamefont{Glasserman,
  Heidelberger, and Shahabuddin}}]{Glasserman2000}
\bibinfo{author}{\bibfnamefont{P.}~\bibnamefont{Glasserman}},
  \bibinfo{author}{\bibfnamefont{P.}~\bibnamefont{Heidelberger}},
  \bibnamefont{and}
  \bibinfo{author}{\bibfnamefont{P.}~\bibnamefont{Shahabuddin}}, in
  \emph{\bibinfo{booktitle}{Mastering Risk}} (\bibinfo{year}{2000}),
  vol.~\bibinfo{volume}{2}, pp. \bibinfo{pages}{5--18}.

\bibitem[{\citenamefont{{Basel Committee on Banking
  Supervision}}(2009)}]{BaselIII}
\bibinfo{author}{\bibnamefont{{Basel Committee on Banking Supervision}}},
  \emph{\bibinfo{title}{Revisions to the basel ii market risk framework}}
  (\bibinfo{year}{2009}).

\bibitem[{\citenamefont{Garc\'ia Ca\~nizares and
  Gen\c{c}ay}(2006)}]{Garcia2006}
\bibinfo{author}{\bibfnamefont{A.}~\bibnamefont{Garc\'ia Ca\~nizares}}
  \bibnamefont{and}
  \bibinfo{author}{\bibfnamefont{R.}~\bibnamefont{Gen\c{c}ay}},
  \bibinfo{journal}{SSRN Electronic Journal}  (\bibinfo{year}{2006}).

\bibitem[{\citenamefont{Nielsen and Chuang}(2010)}]{Nielsen2010}
\bibinfo{author}{\bibfnamefont{M.~A.} \bibnamefont{Nielsen}} \bibnamefont{and}
  \bibinfo{author}{\bibfnamefont{I.~L.} \bibnamefont{Chuang}},
  \emph{\bibinfo{title}{{Quantum Computation and Quantum Information}}}
  (\bibinfo{year}{2010}), ISBN \bibinfo{isbn}{9780511976667}.

\bibitem[{\citenamefont{Peruzzo et~al.}(2014)\citenamefont{Peruzzo, McClean,
  Shadbolt, Yung, Zhou, Love, Aspuru-Guzik, and O'Brien}}]{Peruzzo2014}
\bibinfo{author}{\bibfnamefont{A.}~\bibnamefont{Peruzzo}},
  \bibinfo{author}{\bibfnamefont{J.}~\bibnamefont{McClean}},
  \bibinfo{author}{\bibfnamefont{P.}~\bibnamefont{Shadbolt}},
  \bibinfo{author}{\bibfnamefont{M.~H.} \bibnamefont{Yung}},
  \bibinfo{author}{\bibfnamefont{X.~Q.} \bibnamefont{Zhou}},
  \bibinfo{author}{\bibfnamefont{P.~J.} \bibnamefont{Love}},
  \bibinfo{author}{\bibfnamefont{A.}~\bibnamefont{Aspuru-Guzik}},
  \bibnamefont{and} \bibinfo{author}{\bibfnamefont{J.~L.}
  \bibnamefont{O'Brien}}, \bibinfo{journal}{Nature Communications}
  \textbf{\bibinfo{volume}{5}} (\bibinfo{year}{2014}), ISSN
  \bibinfo{issn}{20411723}.

\bibitem[{\citenamefont{Moll et~al.}(2017)\citenamefont{Moll, Barkoutsos,
  Bishop, Chow, Cross, Egger, Filipp, Fuhrer, Gambetta, Ganzhorn
  et~al.}}]{Moll2017}
\bibinfo{author}{\bibfnamefont{N.}~\bibnamefont{Moll}},
  \bibinfo{author}{\bibfnamefont{P.}~\bibnamefont{Barkoutsos}},
  \bibinfo{author}{\bibfnamefont{L.~S.} \bibnamefont{Bishop}},
  \bibinfo{author}{\bibfnamefont{J.~M.} \bibnamefont{Chow}},
  \bibinfo{author}{\bibfnamefont{A.}~\bibnamefont{Cross}},
  \bibinfo{author}{\bibfnamefont{D.~J.} \bibnamefont{Egger}},
  \bibinfo{author}{\bibfnamefont{S.}~\bibnamefont{Filipp}},
  \bibinfo{author}{\bibfnamefont{A.}~\bibnamefont{Fuhrer}},
  \bibinfo{author}{\bibfnamefont{J.~M.} \bibnamefont{Gambetta}},
  \bibinfo{author}{\bibfnamefont{M.}~\bibnamefont{Ganzhorn}},
  \bibnamefont{et~al.}, \bibinfo{journal}{arXiv:1710.01022}
  (\bibinfo{year}{2017}).

\bibitem[{\citenamefont{Biamonte et~al.}(2017)\citenamefont{Biamonte, Wittek,
  Pancotti, Rebentrost, Wiebe, and Lloyd}}]{Biamonte2017}
\bibinfo{author}{\bibfnamefont{J.}~\bibnamefont{Biamonte}},
  \bibinfo{author}{\bibfnamefont{P.}~\bibnamefont{Wittek}},
  \bibinfo{author}{\bibfnamefont{N.}~\bibnamefont{Pancotti}},
  \bibinfo{author}{\bibfnamefont{P.}~\bibnamefont{Rebentrost}},
  \bibinfo{author}{\bibfnamefont{N.}~\bibnamefont{Wiebe}}, \bibnamefont{and}
  \bibinfo{author}{\bibfnamefont{S.}~\bibnamefont{Lloyd}},
  \bibinfo{journal}{Nature} \textbf{\bibinfo{volume}{549}},
  \bibinfo{pages}{195} (\bibinfo{year}{2017}), ISSN \bibinfo{issn}{14764687}.

\bibitem[{\citenamefont{Brassard et~al.}(2000)\citenamefont{Brassard, Hoyer,
  Mosca, and Tapp}}]{Brassard2000}
\bibinfo{author}{\bibfnamefont{G.}~\bibnamefont{Brassard}},
  \bibinfo{author}{\bibfnamefont{P.}~\bibnamefont{Hoyer}},
  \bibinfo{author}{\bibfnamefont{M.}~\bibnamefont{Mosca}}, \bibnamefont{and}
  \bibinfo{author}{\bibfnamefont{A.}~\bibnamefont{Tapp}},
  \bibinfo{journal}{arXiv:0005055}  (\bibinfo{year}{2000}).

\bibitem[{\citenamefont{Black and Scholes}(1973)}]{BlackScholes}
\bibinfo{author}{\bibfnamefont{F.}~\bibnamefont{Black}} \bibnamefont{and}
  \bibinfo{author}{\bibfnamefont{M.}~\bibnamefont{Scholes}},
  \bibinfo{journal}{Journal of Political Economy}
  \textbf{\bibinfo{volume}{81}}, \bibinfo{pages}{637} (\bibinfo{year}{1973}).

\bibitem[{\citenamefont{Rebentrost et~al.}(2018)\citenamefont{Rebentrost, Gupt,
  and Bromley}}]{Rebentrost2018}
\bibinfo{author}{\bibfnamefont{P.}~\bibnamefont{Rebentrost}},
  \bibinfo{author}{\bibfnamefont{B.}~\bibnamefont{Gupt}}, \bibnamefont{and}
  \bibinfo{author}{\bibfnamefont{T.~R.} \bibnamefont{Bromley}}
  (\bibinfo{year}{2018}), \eprint{arXiv:1805.00109}.

\bibitem[{\citenamefont{{The Bureau of the fiscal
  service}}(2017)}]{PublicDebt201612}
\bibinfo{author}{\bibnamefont{{The Bureau of the fiscal service}}},
  \emph{\bibinfo{title}{Monthly statment of the public debt of the united
  states}} (\bibinfo{year}{2017}).

\bibitem[{\citenamefont{{Federal reserve bank of New
  York}}(2016)}]{TreasuryVolume2016}
\bibinfo{author}{\bibnamefont{{Federal reserve bank of New York}}},
  \emph{\bibinfo{title}{{US Treasury Trading Volume}}} (\bibinfo{year}{2016}).

\bibitem[{\citenamefont{{Federal Reserve Bank of New
  York}}(2018)}]{TreasuryCollateral}
\bibinfo{author}{\bibnamefont{{Federal Reserve Bank of New York}}},
  \emph{\bibinfo{title}{Federal reserve collateral guidelines}}
  (\bibinfo{year}{2018}).

\bibitem[{\citenamefont{Kitaev}(1995)}]{Kitaev1995}
\bibinfo{author}{\bibfnamefont{A.~Y.} \bibnamefont{Kitaev}}
  (\bibinfo{year}{1995}), \eprint{arXiv:9511026}.

\bibitem[{\citenamefont{Abrams and Williams}(1999)}]{Abrams1999}
\bibinfo{author}{\bibfnamefont{D.~S.} \bibnamefont{Abrams}} \bibnamefont{and}
  \bibinfo{author}{\bibfnamefont{C.~P.} \bibnamefont{Williams}}
  (\bibinfo{year}{1999}), \eprint{arXiv:9908083}.

\bibitem[{\citenamefont{Montanaro}(2017)}]{Montanaro2017}
\bibinfo{author}{\bibfnamefont{A.}~\bibnamefont{Montanaro}}
  (\bibinfo{year}{2017}), \eprint{arXiv:1504.06987}.

\bibitem[{\citenamefont{Mottonen et~al.}(2004)\citenamefont{Mottonen,
  Vartiainen, Bergholm, and Salomaa}}]{Mottonen2004a}
\bibinfo{author}{\bibfnamefont{M.}~\bibnamefont{Mottonen}},
  \bibinfo{author}{\bibfnamefont{J.~J.} \bibnamefont{Vartiainen}},
  \bibinfo{author}{\bibfnamefont{V.}~\bibnamefont{Bergholm}}, \bibnamefont{and}
  \bibinfo{author}{\bibfnamefont{M.~M.} \bibnamefont{Salomaa}},
  \bibinfo{journal}{Quant. Inf. Comp.} \textbf{\bibinfo{volume}{5}},
  \bibinfo{pages}{4} (\bibinfo{year}{2004}), ISSN \bibinfo{issn}{15337146},
  \eprint{0407010}.

\bibitem[{\citenamefont{Grover and Rudolph}(2002)}]{Grover2002}
\bibinfo{author}{\bibfnamefont{L.}~\bibnamefont{Grover}} \bibnamefont{and}
  \bibinfo{author}{\bibfnamefont{T.}~\bibnamefont{Rudolph}}
  (\bibinfo{year}{2002}), \eprint{arXiv:0208112}.

\bibitem[{\citenamefont{Vedral et~al.}(1995)\citenamefont{Vedral, Barenco, and
  Ekert}}]{Vedral1995}
\bibinfo{author}{\bibfnamefont{V.}~\bibnamefont{Vedral}},
  \bibinfo{author}{\bibfnamefont{A.}~\bibnamefont{Barenco}}, \bibnamefont{and}
  \bibinfo{author}{\bibfnamefont{A.}~\bibnamefont{Ekert}}
  (\bibinfo{year}{1995}), \eprint{arXiv:9511018v1}.

\bibitem[{\citenamefont{Draper}(2000)}]{Draper2000}
\bibinfo{author}{\bibfnamefont{T.~G.} \bibnamefont{Draper}}
  (\bibinfo{year}{2000}), ISSN \bibinfo{issn}{16136829},
  \eprint{arXiv:0008033}.

\bibitem[{\citenamefont{Cuccaro et~al.}(2004)\citenamefont{Cuccaro, Draper,
  Kutin, and Moulton}}]{Cuccaro2004}
\bibinfo{author}{\bibfnamefont{S.~A.} \bibnamefont{Cuccaro}},
  \bibinfo{author}{\bibfnamefont{T.~G.} \bibnamefont{Draper}},
  \bibinfo{author}{\bibfnamefont{S.~A.} \bibnamefont{Kutin}}, \bibnamefont{and}
  \bibinfo{author}{\bibfnamefont{D.~P.} \bibnamefont{Moulton}}
  (\bibinfo{year}{2004}), \eprint{arXiv:0410184}.

\bibitem[{\citenamefont{Draper et~al.}(2004)\citenamefont{Draper, Kutin, Rains,
  and Svore}}]{Draper2004}
\bibinfo{author}{\bibfnamefont{T.~G.} \bibnamefont{Draper}},
  \bibinfo{author}{\bibfnamefont{S.~A.} \bibnamefont{Kutin}},
  \bibinfo{author}{\bibfnamefont{E.~M.} \bibnamefont{Rains}}, \bibnamefont{and}
  \bibinfo{author}{\bibfnamefont{K.~M.} \bibnamefont{Svore}}, pp.
  \bibinfo{pages}{1--21} (\bibinfo{year}{2004}), ISSN
  \bibinfo{issn}{1533-7146}, \eprint{arXiv:0406142}.

\bibitem[{\citenamefont{Bhaskar et~al.}(2015)\citenamefont{Bhaskar, Hadfield,
  Papageorgiou, and Petras}}]{Bhaskar2015}
\bibinfo{author}{\bibfnamefont{M.~K.} \bibnamefont{Bhaskar}},
  \bibinfo{author}{\bibfnamefont{S.}~\bibnamefont{Hadfield}},
  \bibinfo{author}{\bibfnamefont{A.}~\bibnamefont{Papageorgiou}},
  \bibnamefont{and} \bibinfo{author}{\bibfnamefont{I.}~\bibnamefont{Petras}}
  (\bibinfo{year}{2015}), ISSN \bibinfo{issn}{15337146},
  \eprint{arXiv:1511.08253}.

\bibitem[{\citenamefont{Green et~al.}(2013{\natexlab{a}})\citenamefont{Green,
  Lumsdaine, Ross, Selinger, and Valiron}}]{Green2013}
\bibinfo{author}{\bibfnamefont{A.~S.} \bibnamefont{Green}},
  \bibinfo{author}{\bibfnamefont{P.~L.} \bibnamefont{Lumsdaine}},
  \bibinfo{author}{\bibfnamefont{N.~J.} \bibnamefont{Ross}},
  \bibinfo{author}{\bibfnamefont{P.}~\bibnamefont{Selinger}}, \bibnamefont{and}
  \bibinfo{author}{\bibfnamefont{B.}~\bibnamefont{Valiron}}
  (\bibinfo{year}{2013}{\natexlab{a}}), \eprint{arXiv:1304.3390}.

\bibitem[{\citenamefont{Green et~al.}(2013{\natexlab{b}})\citenamefont{Green,
  Lumsdaine, Ross, Selinger, and Valiron}}]{Green2013b}
\bibinfo{author}{\bibfnamefont{A.~S.} \bibnamefont{Green}},
  \bibinfo{author}{\bibfnamefont{P.~L.} \bibnamefont{Lumsdaine}},
  \bibinfo{author}{\bibfnamefont{N.~J.} \bibnamefont{Ross}},
  \bibinfo{author}{\bibfnamefont{P.}~\bibnamefont{Selinger}}, \bibnamefont{and}
  \bibinfo{author}{\bibfnamefont{B.}~\bibnamefont{Valiron}}, in
  \emph{\bibinfo{booktitle}{Lecture Notes in Computer Science (including
  subseries Lecture Notes in Artificial Intelligence and Lecture Notes in
  Bioinformatics)}} (\bibinfo{year}{2013}{\natexlab{b}}), vol.
  \bibinfo{volume}{7948 LNCS}, pp. \bibinfo{pages}{110--124}, ISBN
  \bibinfo{isbn}{9783642389856}, ISSN \bibinfo{issn}{03029743},
  \eprint{arXiv:1304.5485}.

\bibitem[{\citenamefont{Mitarai et~al.}(2018)\citenamefont{Mitarai, Kitagawa,
  and Fujii}}]{Mitarai2018}
\bibinfo{author}{\bibfnamefont{K.}~\bibnamefont{Mitarai}},
  \bibinfo{author}{\bibfnamefont{M.}~\bibnamefont{Kitagawa}}, \bibnamefont{and}
  \bibinfo{author}{\bibfnamefont{K.}~\bibnamefont{Fujii}}
  (\bibinfo{year}{2018}), \eprint{arXiv:1805.11250}.

\bibitem[{\citenamefont{H{\"{a}}ner et~al.}(2018)\citenamefont{H{\"{a}}ner,
  Roetteler, and Svore}}]{Haner2018}
\bibinfo{author}{\bibfnamefont{T.}~\bibnamefont{H{\"{a}}ner}},
  \bibinfo{author}{\bibfnamefont{M.}~\bibnamefont{Roetteler}},
  \bibnamefont{and} \bibinfo{author}{\bibfnamefont{K.~M.} \bibnamefont{Svore}}
  (\bibinfo{year}{2018}), \eprint{arXiv:1805.12445}.

\bibitem[{\citenamefont{Barenco et~al.}(1995)\citenamefont{Barenco, Bennett,
  Cleve, Divincenzo, Margolus, Shor, Sleator, Smolin, and
  Weinfurter}}]{Barenco1995}
\bibinfo{author}{\bibfnamefont{A.}~\bibnamefont{Barenco}},
  \bibinfo{author}{\bibfnamefont{C.~H.} \bibnamefont{Bennett}},
  \bibinfo{author}{\bibfnamefont{R.}~\bibnamefont{Cleve}},
  \bibinfo{author}{\bibfnamefont{D.~P.} \bibnamefont{Divincenzo}},
  \bibinfo{author}{\bibfnamefont{N.}~\bibnamefont{Margolus}},
  \bibinfo{author}{\bibfnamefont{P.}~\bibnamefont{Shor}},
  \bibinfo{author}{\bibfnamefont{T.}~\bibnamefont{Sleator}},
  \bibinfo{author}{\bibfnamefont{J.~A.} \bibnamefont{Smolin}},
  \bibnamefont{and}
  \bibinfo{author}{\bibfnamefont{H.}~\bibnamefont{Weinfurter}},
  \bibinfo{journal}{Physical Review A} \textbf{\bibinfo{volume}{52}},
  \bibinfo{pages}{3457} (\bibinfo{year}{1995}), ISSN \bibinfo{issn}{10502947},
  \eprint{arXiv:9503016}.

\bibitem[{\citenamefont{Trefethen}(2013)}]{Trefethen}
\bibinfo{author}{\bibfnamefont{L.~N. L.~N.} \bibnamefont{Trefethen}},
  \emph{\bibinfo{title}{{Approximation theory and approximation practice}}}
  (\bibinfo{year}{2013}), ISBN \bibinfo{isbn}{1611972396}.

\bibitem[{\citenamefont{Black et~al.}(1990)\citenamefont{Black, Derman, and
  Toy}}]{black1990}
\bibinfo{author}{\bibfnamefont{F.}~\bibnamefont{Black}},
  \bibinfo{author}{\bibfnamefont{E.}~\bibnamefont{Derman}}, \bibnamefont{and}
  \bibinfo{author}{\bibfnamefont{W.}~\bibnamefont{Toy}},
  \bibinfo{journal}{Financial Analysts Journal} \textbf{\bibinfo{volume}{46}},
  \bibinfo{pages}{33} (\bibinfo{year}{1990}), ISSN \bibinfo{issn}{0015198X}.

\bibitem[{ibm({\natexlab{a}})}]{ibm_q_yorktown_processor}
\emph{\bibinfo{title}{{IBM Q 5 (ibmqx2)}}},
  \bibinfo{howpublished}{\url{https://github.com/QISKit/qiskit-backend-information/blob/master/backends/yorktown/V1/README.md}},
  \bibinfo{note}{accessed: 2018-05-22}.

\bibitem[{\citenamefont{Nippani et~al.}(2001)\citenamefont{Nippani, Liu, and
  Schulman}}]{Nippani2001}
\bibinfo{author}{\bibfnamefont{S.}~\bibnamefont{Nippani}},
  \bibinfo{author}{\bibfnamefont{P.}~\bibnamefont{Liu}}, \bibnamefont{and}
  \bibinfo{author}{\bibfnamefont{C.~T.} \bibnamefont{Schulman}},
  \bibinfo{journal}{The Journal of Financial and Quantitative Analysis}
  \textbf{\bibinfo{volume}{36}}, \bibinfo{pages}{251} (\bibinfo{year}{2001}).

\bibitem[{\citenamefont{Colin et~al.}(2006)\citenamefont{Colin, Cubili\'e, and
  Bardoux}}]{Colin2006}
\bibinfo{author}{\bibfnamefont{A.}~\bibnamefont{Colin}},
  \bibinfo{author}{\bibfnamefont{M.}~\bibnamefont{Cubili\'e}},
  \bibnamefont{and} \bibinfo{author}{\bibfnamefont{F.}~\bibnamefont{Bardoux}},
  \bibinfo{journal}{The Journal of Performance Measurement}
  \textbf{\bibinfo{volume}{10}} (\bibinfo{year}{2006}).

\bibitem[{\citenamefont{Vannerem and Iyer}(2010)}]{Vannerem2010}
\bibinfo{author}{\bibfnamefont{P.}~\bibnamefont{Vannerem}} \bibnamefont{and}
  \bibinfo{author}{\bibfnamefont{A.~S.} \bibnamefont{Iyer}},
  \bibinfo{journal}{MSCI Barra Research Insights}  (\bibinfo{year}{2010}).

\bibitem[{\citenamefont{Martellini et~al.}(2003)\citenamefont{Martellini,
  Priaulet, and Priaulet}}]{Martellini2003}
\bibinfo{author}{\bibfnamefont{L.}~\bibnamefont{Martellini}},
  \bibinfo{author}{\bibfnamefont{P.}~\bibnamefont{Priaulet}}, \bibnamefont{and}
  \bibinfo{author}{\bibfnamefont{S.}~\bibnamefont{Priaulet}},
  \emph{\bibinfo{title}{Fixed-Income Securities}} (\bibinfo{publisher}{Wiley
  and Sons}, \bibinfo{year}{2003}), ISBN \bibinfo{isbn}{0-470-85277-1}.

\bibitem[{\citenamefont{Sheldon et~al.}(2016)\citenamefont{Sheldon, Magesan,
  Chow, and Gambetta}}]{Sheldon2016b}
\bibinfo{author}{\bibfnamefont{S.}~\bibnamefont{Sheldon}},
  \bibinfo{author}{\bibfnamefont{E.}~\bibnamefont{Magesan}},
  \bibinfo{author}{\bibfnamefont{J.~M.} \bibnamefont{Chow}}, \bibnamefont{and}
  \bibinfo{author}{\bibfnamefont{J.~M.} \bibnamefont{Gambetta}},
  \bibinfo{journal}{Phys. Rev. A} \textbf{\bibinfo{volume}{93}},
  \bibinfo{pages}{060302} (\bibinfo{year}{2016}).

\bibitem[{\citenamefont{Gustavsson et~al.}(2013)\citenamefont{Gustavsson,
  Zwier, Bylander, Yan, Yoshihara, Nakamura, Orlando, and
  Oliver}}]{Gustavsson2013}
\bibinfo{author}{\bibfnamefont{S.}~\bibnamefont{Gustavsson}},
  \bibinfo{author}{\bibfnamefont{O.}~\bibnamefont{Zwier}},
  \bibinfo{author}{\bibfnamefont{J.}~\bibnamefont{Bylander}},
  \bibinfo{author}{\bibfnamefont{F.}~\bibnamefont{Yan}},
  \bibinfo{author}{\bibfnamefont{F.}~\bibnamefont{Yoshihara}},
  \bibinfo{author}{\bibfnamefont{Y.}~\bibnamefont{Nakamura}},
  \bibinfo{author}{\bibfnamefont{T.~P.} \bibnamefont{Orlando}},
  \bibnamefont{and} \bibinfo{author}{\bibfnamefont{W.~D.}
  \bibnamefont{Oliver}}, \bibinfo{journal}{Phys. Rev. Lett.}
  \textbf{\bibinfo{volume}{110}}, \bibinfo{pages}{040502}
  (\bibinfo{year}{2013}).

\bibitem[{\citenamefont{{International Business Machines
  Corporation}}(2016)}]{QuantumExperience}
\bibinfo{author}{\bibnamefont{{International Business Machines Corporation}}},
  \emph{\bibinfo{title}{{IBM Q Experience}}} (\bibinfo{year}{2016}),
  \urlprefix\url{https://quantumexperience.ng.bluemix.net/qx/experience}.

\bibitem[{Qis(2018)}]{Qiskit}
\emph{\bibinfo{title}{{Quantum Information Software Kit (QISKit)}}}
  (\bibinfo{year}{2018}), \urlprefix\url{https://qiskit.org/}.

\bibitem[{\citenamefont{Rigetti and Devoret}(2010)}]{Rigetti2010}
\bibinfo{author}{\bibfnamefont{C.}~\bibnamefont{Rigetti}} \bibnamefont{and}
  \bibinfo{author}{\bibfnamefont{M.}~\bibnamefont{Devoret}},
  \bibinfo{journal}{Phys. Rev. B} \textbf{\bibinfo{volume}{81}},
  \bibinfo{pages}{134507} (\bibinfo{year}{2010}).

\bibitem[{\citenamefont{Dobsicek et~al.}(2006)\citenamefont{Dobsicek,
  Johansson, Shumeiko, and Wendin}}]{Dobsicek2006}
\bibinfo{author}{\bibfnamefont{M.}~\bibnamefont{Dobsicek}},
  \bibinfo{author}{\bibfnamefont{G.}~\bibnamefont{Johansson}},
  \bibinfo{author}{\bibfnamefont{V.}~\bibnamefont{Shumeiko}}, \bibnamefont{and}
  \bibinfo{author}{\bibfnamefont{G.}~\bibnamefont{Wendin}},
  \bibinfo{journal}{Science} \textbf{\bibinfo{volume}{2}}, \bibinfo{pages}{2}
  (\bibinfo{year}{2006}), \eprint{arXiv:0610214}.

\bibitem[{\citenamefont{O'Loan}(2010)}]{OLoan2010}
\bibinfo{author}{\bibfnamefont{C.~J.} \bibnamefont{O'Loan}},
  \bibinfo{journal}{Journal of Physics A: Mathematical and Theoretical}
  \textbf{\bibinfo{volume}{43}} (\bibinfo{year}{2010}), ISSN
  \bibinfo{issn}{17518113}, \eprint{arXiv:0904.3426}.

\bibitem[{\citenamefont{Svore et~al.}(2014)\citenamefont{Svore, Hastings, and
  Freedman}}]{Svore2013}
\bibinfo{author}{\bibfnamefont{K.~M.} \bibnamefont{Svore}},
  \bibinfo{author}{\bibfnamefont{M.~B.} \bibnamefont{Hastings}},
  \bibnamefont{and} \bibinfo{author}{\bibfnamefont{M.}~\bibnamefont{Freedman}},
  \bibinfo{journal}{Quantum Information {\&} Computation}
  \textbf{\bibinfo{volume}{14}}, \bibinfo{pages}{306} (\bibinfo{year}{2014}),
  ISSN \bibinfo{issn}{1533-7146}, \eprint{arXiv:1304.0741v1}.

\bibitem[{\citenamefont{Egger et~al.}(2018)\citenamefont{Egger, Ganzhorn,
  Fuhrer, Mueler, Barkoutsos, Moll, Tavernelli, and Filipp}}]{Egger2018b}
\bibinfo{author}{\bibfnamefont{D.~J.} \bibnamefont{Egger}},
  \bibinfo{author}{\bibfnamefont{G.}~\bibnamefont{Ganzhorn},
  \bibfnamefont{Marc~Salis}},
  \bibinfo{author}{\bibfnamefont{A.}~\bibnamefont{Fuhrer}},
  \bibinfo{author}{\bibfnamefont{P.}~\bibnamefont{Mueler}},
  \bibinfo{author}{\bibfnamefont{P.~K.} \bibnamefont{Barkoutsos}},
  \bibinfo{author}{\bibfnamefont{N.}~\bibnamefont{Moll}},
  \bibinfo{author}{\bibfnamefont{I.}~\bibnamefont{Tavernelli}},
  \bibnamefont{and} \bibinfo{author}{\bibfnamefont{S.}~\bibnamefont{Filipp}}
  (\bibinfo{year}{2018}), \eprint{arXiv:1804.04900}.

\bibitem[{\citenamefont{Sj\"oqvist et~al.}(2012)\citenamefont{Sj\"oqvist, Tong,
  Andersson, Hessmo, Johansson, and Singh}}]{Sjoqvist2012}
\bibinfo{author}{\bibfnamefont{E.}~\bibnamefont{Sj\"oqvist}},
  \bibinfo{author}{\bibfnamefont{D.~M.} \bibnamefont{Tong}},
  \bibinfo{author}{\bibfnamefont{L.~M.} \bibnamefont{Andersson}},
  \bibinfo{author}{\bibfnamefont{B.}~\bibnamefont{Hessmo}},
  \bibinfo{author}{\bibfnamefont{M.}~\bibnamefont{Johansson}},
  \bibnamefont{and} \bibinfo{author}{\bibfnamefont{K.}~\bibnamefont{Singh}},
  \bibinfo{journal}{New J. Phys.} \textbf{\bibinfo{volume}{14}}
  (\bibinfo{year}{2012}).

\bibitem[{\citenamefont{Kandala et~al.}(2018)\citenamefont{Kandala, Temme,
  D.~Corcoles, Mezzacapo, Chow, and Gambetta}}]{Kandala2018}
\bibinfo{author}{\bibfnamefont{A.}~\bibnamefont{Kandala}},
  \bibinfo{author}{\bibfnamefont{K.}~\bibnamefont{Temme}},
  \bibinfo{author}{\bibfnamefont{A.}~\bibnamefont{D.~Corcoles}},
  \bibinfo{author}{\bibfnamefont{A.}~\bibnamefont{Mezzacapo}},
  \bibinfo{author}{\bibfnamefont{J.~M.} \bibnamefont{Chow}}, \bibnamefont{and}
  \bibinfo{author}{\bibfnamefont{J.~M.} \bibnamefont{Gambetta}}
  (\bibinfo{year}{2018}), \eprint{arXiv:1805.04492}.

\bibitem[{\citenamefont{Tokdar and Kass}(2010)}]{Tokdar2010}
\bibinfo{author}{\bibfnamefont{S.~T.} \bibnamefont{Tokdar}} \bibnamefont{and}
  \bibinfo{author}{\bibfnamefont{R.~E.} \bibnamefont{Kass}},
  \bibinfo{journal}{Wiley Interdisciplinary Reviews: Computational Statistics}
  \textbf{\bibinfo{volume}{2}}, \bibinfo{pages}{54} (\bibinfo{year}{2010}),
  ISSN \bibinfo{issn}{19395108}.

\bibitem[{\citenamefont{{U.S. Department of the Treasury}}(2018)}]{treasury}
\bibinfo{author}{\bibnamefont{{U.S. Department of the Treasury}}},
  \emph{\bibinfo{title}{Daily treasury yield curve rates}}
  (\bibinfo{year}{2018}).

\bibitem[{ibm({\natexlab{b}})}]{ibm_q_austin_processor}
\emph{\bibinfo{title}{{IBM Q 20}}},
  \bibinfo{howpublished}{\url{https://quantumexperience.ng.bluemix.net/qx/devices}},
  \bibinfo{note}{accessed: 2018-05-22}.

\end{thebibliography}

\end{document}